\documentclass{JHEP3}
\pdfoutput=1
\usepackage{amsmath,amssymb,epsfig}

\newcommand{\be}{\begin{equation}}
\newcommand{\ee}{\end{equation}}
\newcommand{\beq}{\begin{equation}}
\newcommand{\eeq}{\end{equation}}
\newcommand{\ba}{\begin{eqnarray}}
\newcommand{\ea}{\end{eqnarray}}
\newcommand{\bea}{\begin{eqnarray}}
\newcommand{\eea}{\end{eqnarray}}

\def\IZ {\mathbb{Z}}

\def\vt#1#2#3 {{\vartheta[{#1 \atop  #2}](#3\vert \tau)}}

\title{Causality constraints in AdS/CFT from conformal collider physics and Gauss-Bonnet gravity}
\author{Xi\'an O. Camanho\\
\sl Department of Particle Physics and IGFAE, University of Santiago de Compostela, E-15782 Santiago de Compostela, Spain\\\vskip-4mm
\email{xian.otero@rai.usc.es}}

\author{Jos\'e D. Edelstein\\
\sl Department of Particle Physics and IGFAE, University of Santiago de Compostela, E-15782 Santiago de Compostela, Spain\\\vskip-3mm
\sl Centro de Estudios Cient\'\i ficos, Valdivia, Chile\\\vskip-4mm
\email{jose.edelstein@usc.es}}

\bigskip
\bigskip
\abstract{We explore the relation between positivity of the energy constraints in conformal field theories and causality in their dual gravity description. Our discussion involves CFTs with different central charges whose description, in the gravity side, requires the inclusion of quadratic curvature corrections. It is enough, indeed, to consider the Gauss-Bonnet term. We find that both sides of the AdS/CFT correspondence impose a restriction on the Gauss-Bonnet coupling. In the case of 6d supersymmetric CFTs, we show the full matching of these restrictions. We perform this computation in two ways. First by considering a thermal setup in a black hole background. Second by scrutinizing the scattering of gravitons with a shock wave in AdS. The different helicities provide the corresponding lower and upper bounds. We generalize these results to arbitrary higher dimensions and comment on some hints and puzzles they prompt regarding the possible existence of higher dimensional CFTs and the extent to which the AdS/CFT correspondence would be valid for them.}

\preprint{arXiv:0911.3160 [hep-th]}

\begin{document}

%%%%%%%%%%%%%%%%%%%%%%%%%%%%%%%%%%%%%%%%%%%%%%%%%%%%
\section{Introduction}
%%%%%%%%%%%%%%%%%%%%%%%%%%%%%%%%%%%%%%%%%%%%%%%%%%%%

The AdS/CFT correspondence is by now a well-established non-perturbative duality of paramount importance. After several years of research, it has overpassed dozens of checks and has been applied in a plethora of systems that go far beyond the large $N$ limit of $\mathcal{N} = 4$ super Yang-Mills theory in four dimensions, originally portrayed by Maldacena \cite{Maldacena1998}. Some of the most convincing tests have been recently reviewed in \cite{Klebanov2008a}.

Despite of this accumulating evidence, a novel direction was recently explored by Hofman and Maldacena \cite{Hofman2008}. Based on a framework developed earlier for e$^+$-- e$^-$ annihilation in QCD \cite{Basham1978,Basham1978a}, they studied a {\it gedanken} collider physics setup in the context of conformal field theories. They focused in the case of 4d CFTs, and found a number of constraints for the central charges by demanding that the energy measured in calorimeters of a collider physics experiment be positive. They found, for instance, that any 4d $\mathcal{N}=1$ supersymmetric CFT must have central charges\footnote{The conformal anomaly of a four-dimensional CFT can be obtained by computing the trace of the stress-energy tensor in a curved spacetime \cite{Birrell}
\begin{equation}
\langle T^\mu{}_\mu \rangle_{\rm CFT} = \frac{c}{16\pi^2} I_4 - \frac{a}{16\pi^2} E_4 ~,
\label{vevTmunu}
\end{equation}
where $c$ and $a$ are the central charges, and $E_4$ and $I_4$ correspond to the four-dimensional Euler density and the square of the Weyl curvature.} within the window, $1/2 \leq a/c \leq 3/2$, the bounds being saturated by free theories with only chiral supermultiplets (lower bound) or only vector supermultiplets (upper bound) \cite{Hofman2008}. Since the computation of $\langle T^\mu_{~\mu} \rangle$ in a state generated by the stress-energy tensor is given by 3-point correlators of $T$, and pure Einstein-Hilbert gravity is well-known to yield $a = c$ \cite{Henningson1998,Henningson2000}, the gravity dual of a theory with $a \neq c$ should contain higher (at least quadratic) curvature corrections.

In a seemingly different context, Brigante {\it et al.} \cite{Brigante2008,Brigante2008a} explored the addition of a Gauss-Bonnet term in the gravity side of the AdS/CFT correspondence and showed that, in the background of a black hole, the coefficient of this term, $\lambda$, is bounded from above, $\lambda \leq 9/100$, in order to preserve causality at the boundary.\footnote{Indeed, this is more general since any other curvature squared term can be reduced to Gauss-Bonnet by field redefinitions disregarding higher powers of the curvature. See for instance \cite{Brigante2008}. Some related work along these lines has been pursued in \cite{Neupane2009a,Neupane2009f}.} If this bound is disregarded, boundary perturbations would propagate at superluminal velocities. A natural question is immediately raised as for whether this quadratic curvature corrections arise in the string theory framework. The answer was given in the affirmative by Kats and Petrov \cite{Kats2009}, and further explored more recently by Buchel {\it et al.} \cite{Buchel2009}. Both papers focus on string theory compactifications that are relevant in the context of 4d SCFTs.

The somehow striking result came when Hofman and Maldacena realized that the upper bound on $\lambda$ was nothing but, through holographic renormalization, the lower bound on $a/c$. The matching is exact. There is also a lower bound that $\lambda$ has to satisfy due to causality constraints, and it should be possible to rephrase it in terms of the upper bound on $a/c$ \cite{HofMalpvt}. This seems to provide a deep connection between two central concepts such as causality and positivity of the energy in both sides of the AdS/CFT correspondence. Besides, these results provided an irrefutable evidence against the so-called KSS bound \cite{Kovtun2005} for $\eta/s$ in quantum relativistic theories, $\eta/s \geq \frac{1}{4\pi}$. This is due to the fact that the value for $\eta/s$ is corrected in presence of a Gauss-Bonnet correction to $\eta/s = \frac{1}{4\pi} (1 - 4 \lambda)$. Since the upper bound for $\lambda$ is positive, the shear viscosity to entropy density ratio, for such a SCFT, would be lower than the KSS value.

In a recent paper, Buchel and Myers \cite{Buchel2009a} dug further into the constraints imposed by causality in the holographic description of hydrodynamics. Their paper deals with a black hole background, in which they explicitly relate the value of $\lambda$ to the difference between central charges of the dual CFT. They show that a lower bound for the Gauss-Bonnet coupling, $\lambda \geq - 7/36$ comes out due to causality constraints, and that it corresponds precisely to the upper bound $a/c \leq 3/2$.

It was later pointed out by Hofman \cite{Hofman2009}, that bounds resulting from causality constraints should not be a feature of thermal CFTs. The relation between causality and positivity must lie at a more fundamental level and, as such, should show up at zero temperature. Indeed, by means of an ingenuous computation using shock waves, he proved that the upper bound on $\lambda$ comes from causality requirements imposed on a scattering process involving a graviton and the shock wave. It should be pointed out that the positive energy condition in CFTs used in the conformal collider setup is not self-evident at all. Hofman gave a field theoretic argument explaining why it holds in any UV complete quantum field theory \cite{Hofman2009}. He scrutinized deeper in the relation between causality and positive energy, and showed that there are indeed several bounds resulting from the different helicities both of the stress-energy tensor in the CFT side as well as of the metric perturbations in the gravity side.

This perfect match, both qualitative and quantitative, is encouraging and presents new puzzles. The addition of higher curvature corrections in the gravity side has a quantum mechanical nature, thus exploring the holographic principle thoroughly beyond the semiclassical level. It is immediate to ask whether this extends to CFTs in dimensions different than four. A natural candidate to deal with is 6d, since we know that there is a well-studied system in M-theory that corresponds to a $(2,0)$ SCFT in 6d \cite{Witten1995c,Strominger1996d}. If there exists a SCFT in 6d with large central charges but whose difference cannot be neglected in the 't Hooft limit, the gravity dual shall contain terms quadratic in the curvature. This is due to the fact that these differences appear in the 3-point function of the stress-energy tensor and, in the gravity side, this operator is sourced by a 3-graviton vertex.

The relation between causality and positivity of the energy in 6d CFTs was studied very recently by de Boer, Kulaxizi and Parnachev \cite{Boer2009}. These authors courageously performed the holographic renormalization computation that allows to relate the central charges of the CFT with the Gauss-Bonnet coefficient. This case is more complicated than its 4d counterpart since the CFT has three central charges though the positive energy conditions constrain two independent combinations thereof. They studied causality violation in the gravity side and showed that, again, $\lambda$ is bounded from above, $\lambda \leq 3/16$, which further reduces the value of $\eta/s$ in the corresponding plasma. They also showed that this bound is precisely the one arising in the CFT side from positivity of the energy arguments. These latter arguments also lead to a lower bound for the Gauss-Bonnet coupling, $\lambda \geq - 5/16$. The authors of \cite{Boer2009} presume that this lower bound may arise from considering excitations with a different polarization. This is one of the results of the current work.

In this paper we show that the lower bound for $\lambda$ is a causality constraint. We study perturbations of different helicities in the black hole background of Gauss-Bonnet gravity and show that each of them gives a bound that exactly matches the ones arising from positivity of the energy in 6d conformal collider physics. We do this also in the shock wave setup, that gets rid of the notion of a thermal CFT, where we also show that different polarizations of the graviton that collides with a shock wave provide the same one-to-one correspondence to the positivity bounds in the field theory side. These results add up to those previously obtained in \cite{Hofman2009,Boer2009} and altogether they convey a strong piece of evidence supporting the AdS/CFT conjecture.

We generalize all the expressions for an arbitrary higher dimensional AdS/CFT dual pair. This is not, a priori, guaranteed to have any meaning, but it is tempting to explore this possibility and, as we will show, it leads to interesting results. On higher dimensions, though, the holographic renormalization computation of the CFT central charges is missing. Its difficulty increases heavily with space-time dimensionality. It is indeed unclear whether there are non-trivial higher dimensional CFTs. There seems to be an avenue for their formulation in terms of $p$-forms with $p \geq 2$ (so-called generalized gerbe theories) \cite{Witten2007c}. It is still possible to argue within the conformal collider physics setup, on general grounds, that there should be bounds due to positivity of the energy conditions in these conjectural theories. The formulas we obtain for higher $d$ match these expectations. All the expressions are extended smoothly and meaningfully, as we discuss below. This may provide evidence supporting the possibility that AdS/CFT is not necessarily related to string theory.

An interesting puzzle indeed has to do with the string theory origin of quadratic curvature corrections as those of Gauss-Bonnet gravity. These curvature corrections may appear in type II string theory due to $\alpha'$ corrections to the DBI action of probe D-branes \cite{Buchel2009}. The natural context in the 6d case, however, as discussed above, is to see their emergence in M-theory. Even though corrections of this sort are known to exist due to the presence of wrapped M5-branes \cite{Bachas1999}, it is not straighforward to see how they would extend to our case. They will presumably emerge\footnote{We thank Juan Maldacena for his comments about this issue.} from $A_{k-1}$ singularities produced in M-theory by a $\IZ_k$ orbifold of the AdS$_7 \times$ S$^4$ background \cite{Gaiotto2009d}. Indeed, thinking of the $S^4$ as an $S^3$ fibered on $S^1$, modding out by $\IZ_k \subset U(1) \subset SU(2)_{\rm L}$, where $SU(2)_L$ acts on the left on the 3-sphere, after Kaluza-Klein reduction along the $U(1)$ circle, leads to $k$ D6-branes in type IIA string theory (see, for instance, the discussion in \cite{Edelstein2001}). Hence, the $\alpha'$ corrected DBI terms extensively discussed in \cite{Buchel2009} should extend smoothly to our case.

The paper is organized as follows. In Section 2 we introduce some aspects of 6d SCFTs that will be needed later on. A very complete discussion can be found in \cite{Boer2009}. Section 3 is devoted to present the necessary formulas corresponding to Gauss-Bonnet theory in 7d. In Section 4 we analyze boundary perturbations with different helicities of black hole AdS backgrounds, and show how they lead to the different bounds in $\lambda$. Section 5 contains a parallel computation involving the scattering of gravitons and shock waves. We construct the setup for the generic case of pp-waves and reproduce the same bounds in this setup. We complete the discussion initiated in \cite{Hofman2009}, showing explicitly that the lower bounds can be obtained from scalar perturbations. In Section 6 we extend all the results to the case of arbitrary $d$. We obtain matching formulas from the black hole setup and the shock wave. We discuss their implications on the realm of would be higher dimensional strongly coupled CFTs and the AdS/CFT correspondence, and summarize the results, in Section 7.

%%%%%%%%%%%%%%%%%%%%%%%%%%%%%%%%%%%%%%%%%%%%%%%%%%%%
\section{Aspects of 6d CFTs and conformal collider physics}
%%%%%%%%%%%%%%%%%%%%%%%%%%%%%%%%%%%%%%%%%%%%%%%%%%%%

Consider a {\it gedanken} collision experiment in a 6d CFT that mimics the framework developed in \cite{Basham1978,Basham1978a} for e$^+$--e$^-$ annihilation in QCD. We would like to measure the total energy flux per unit angle deposited in calorimeters distributed around the collision region \cite{Hofman2008},
\begin{equation}
{\cal E}(\hat{n}) = \lim_{r \to \infty} r^4\! \int_{-\infty}^\infty\! dt\; n^i\, T^0_{~i}(t, r\, \hat{n}) ~,
\label{etheta}
\end{equation}
the vector $\hat{n}$ pointing towards the actual direction of measure. The expectation value of the energy on a state created by a given local gauge invariant operator $\mathcal{O}$,
\begin{equation}
\langle {\cal E}(\hat{n}) \rangle_\mathcal{O} = \frac{\langle 0| \mathcal{O}^\dagger {\cal E}(\hat{n}) \mathcal{O} |0 \rangle}{\langle 0| \mathcal{O}^\dagger \mathcal{O} |0 \rangle} ~,
\label{vevenergy}
\end{equation}
is written in terms of 2- and 3-point functions in the CFT. There is a natural operator of this sort to be considered, that CFTs in any space-time dimension possesses, which is the stress-energy tensor, $\mathcal{O} = \epsilon_{ij}\, T_{ij}$. For such operators, $\langle {\cal E}(\hat{n}) \rangle_\mathcal{O}$ is given in terms of 2- and 3-point correlators of $T_{\mu\nu}$ \cite{Hofman2008}. These are determined by just a few parameters in any CFT.

In 6d CFTs, in particular, using the fact that $\epsilon_{ij}$ is a symmetric and traceless polarization tensor with purely spatial indices, $O(5)$ rotational symmetry allows to write \cite{Boer2009}
\begin{equation}
\langle {\cal E}(\hat{n}) \rangle_{\epsilon_{ij}\, T_{ij}} = \frac{3\,q_0}{8\,\pi^2} \left[ 1 + t_2 \left( \frac{n_i\,\epsilon^{*}_{il}\,\epsilon_{lj}\,n_j}{\epsilon^{*}_{ij}\,\epsilon_{ij}} - \frac{1}{5} \right) + t_4 \left( \frac{|\epsilon_{ij}\,n_i\,n_j|^2}{\epsilon^{*}_{ij}\,\epsilon_{ij}} - \frac{2}{35} \right) \right] ~,
\label{t2andt4}
\end{equation}
the energy flux being almost completely fixed by symmetry up to coefficients $t_2$ and $t_4$. In a recent paper, de Boer {\it et al.} studied the conditions imposed by demanding positivity of the deposited energy irrespective of the calorimeter angular position. They depend on the different polarizations $\epsilon_{ij}$ \cite{Boer2009},
\begin{eqnarray}
{\rm tensor:}\qquad & & 1 - \frac{1}{5}\, t_2 - \frac{2}{35}\, t_4 \geq 0 ~,\label{tensorbound} \\ [0.7em]
{\rm vector:}\qquad & & \left( 1 - \frac{1}{5}\, t_2 - \frac{2}{35}\, t_4 \right) + \frac{1}{2}\, t_2 \geq 0 ~, \label{vectorbound} \\ [0.7em]
{\rm scalar:}\qquad & & \left( 1 - \frac{1}{5}\, t_2 - \frac{2}{35}\, t_4 \right) + \frac{4}{5} \left( t_2 + t_4 \right) \geq 0 ~. \label{scalarbound}
\end{eqnarray}
The three expressions come from the splitting of $\epsilon_{ij}$ into tensor, vector and scalar components with respect to rotations in the hyperplane perpendicular to $\hat{n}$. These constraints are saturated in a free field theory with, respectively, no antisymmetric tensor fields, no fermions or no scalars \cite{Boer2009}. This is similar to what happens in 4d \cite{Hofman2008}. These constraints restrict the possible values of $t_2$ and $t_4$ for any CFT in 6d to lie inside a triangle (see Figure \ref{triangle}).
%%%%%%%%%%%%%
\FIGURE{\includegraphics[width=0.67\textwidth]{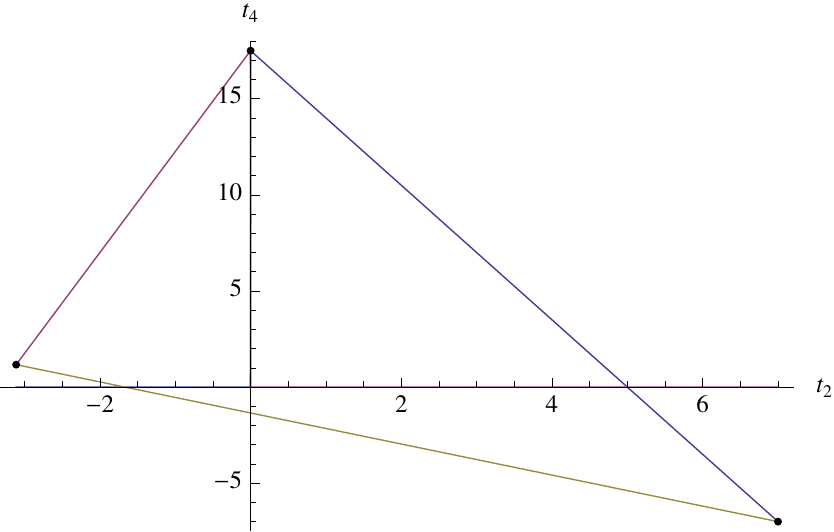}\caption{Constraints (\ref{tensorbound})--(\ref{scalarbound}) restrict the values of $t_2$ and $t_4$ to the interior of the depicted triangle.}
\label{triangle}}
%%%%%%%%%%%%%

The presence of a non-trivial $t_4$ is linked to the absence of supersymmetry \cite{Hofman2008,Boer2009}. In particular, for any supersymmetric CFT in 6d, $t_4$ vanishes and the above constraints translate into bounds that $t_2$ must obey: $t_2 \leq 5$ (tensor), $t_2 \geq - 10/3$ (vector) and $t_2 \geq - 5/3$ (scalar). The restriction imposed by the vector polarization, as it happens in 4d, is less restrictive than the scalar one and, thus, irrelevant. Summarizing, positive energy constraints impose the following restriction on $t_2$ for any 6d SCFT,
\begin{equation}
- \frac{5}{3} \leq t_2 \leq 5 ~.
\label{t2general}
\end{equation}
The 3-point function of the stress-energy tensor in any CFT, irrespective of the space-time dimensionality, can be written in terms of three independent coefficients: $a$, $b$ and $c$ \cite{Osborn1994}. The 2-point function, instead, depends on a unique parameter $C_T$ which is a function of $a$, $b$ and $c$. The ratio, thus, will be determined by two independent parameters (better, two independent combinations of the above coefficients). These should be identified with $t_2$ and $t_4$.\footnote{Notice that this is the case for any space-time dimension. Indeed,
\begin{equation}
\langle {\cal E}(\hat{n}) \rangle_{\epsilon_{ij}\, T_{ij}} = \frac{q_0}{\omega_{d-3}} \left[ 1 + t_2 \left( \frac{n_i\,\epsilon^{*}_{il}\,\epsilon_{lj}\,n_j}{\epsilon^{*}_{ij}\,\epsilon_{ij}} - \frac{1}{d-2} \right) + t_4 \left( \frac{|\epsilon_{ij}\,n_i\,n_j|^2}{\epsilon^{*}_{ij}\,\epsilon_{ij}} - \frac{2}{d (d-2)} \right) \right] ~,
\label{t2andt4-d}
\end{equation}
the $d$ dependence in this expression being given by the normalization of the integrals over $d-3$ sphere whose volume is $\omega_{d-3}$. We always make reference to the dual AdS space-time dimensionality ({\it e.g.}, to obtain (\ref{t2andt4}) one should set $d=7$).} The specific expressions for the latter quantities in terms of the coefficients of the 3-point functions were obtained very recently \cite{Boer2009}. They can also be related to the Weyl anomaly parameters, whose structure in a 6d CFT reads \cite{Bonora1986}
\begin{equation}
\langle T^\mu_{~\mu} \rangle = E_6 + \sum_i^3 c_i\, I_i + \nabla_i J^i ~,
\label{Weylanomaly}
\end{equation}
where $E_6$ is the Euler density (the so-called A-type anomaly), $I_i$ are three independent Weyl invariants of dimension six (B-type anomaly) and $\nabla_i J^i$ is the linear combination of the Weyl variation of six independent local functionals (D-type anomaly). There is a linear relation between the coefficients of the B-type anomaly and $a$, $b$ and $c$ \cite{Boer2009}. This is important since there is a well-defined procedure to compute the Weyl anomaly in AdS/CFT \cite{Henningson1998,Henningson2000}. This allows, at the end of a painful computation that was carried out in \cite{Boer2009}, to obtain the crucial dependence of $t_2$ on the Gauss-Bonnet coupling:
\begin{equation}
t_2 = 5 \left( \frac{1}{\sqrt{1 - 4 \lambda}} - 1 \right) ~.
\label{t2vslambda}
\end{equation}
The above mentioned constraint on $t_2$, given in (\ref{t2general}), translates into a corresponding restriction on $\lambda$,
\begin{equation}
- \frac{5}{16} \leq \lambda \leq \frac{3}{16} ~,
\label{lambda7d}
\end{equation}
the bound imposed by the vector polarization, $\lambda \geq -2$, being less restrictive than that resulting from the scalar polarization. The upper bound was recently shown to be also a consequence of causality restrictions in the gravity side, coming from the analysis of would-be superluminal helicity two perturbations in the boundary of a black hole AdS space-time. We will show below that the lower bound similarly emerges from the analysis of scalar perturbations in the same background.

%%%%%%%%%%%%%%%%%%%%%%%%%%%%%%%%%%%%%%%%%%%%%%%%%%%%
\section{Gauss-Bonnet theory}
%%%%%%%%%%%%%%%%%%%%%%%%%%%%%%%%%%%%%%%%%%%%%%%%%%%%

In this section we review some known facts of the Gauss-Bonnet (GB) theory of gravity. We do this in a quite less familiar formulation that will prove to be very suitable for the purpose of our computations. The standard GB term, in addition to the Einstein-Hilbert and cosmological contributions, give the following action\footnote{We set $16\pi G_N = 1$, without any lose, to simplify our expressions.}
\begin{equation}
\mathcal{I} = \mathop\int d^{d}x \, \sqrt{-g}\, \left[ R - 2 \Lambda + \frac{\lambda\, L^2}{(d-3)(d-4)} \left( R^2 - 4R_{\mu\nu} R^{\mu\nu} + R_{\mu\nu\rho\sigma} R^{\mu\nu\rho\sigma} \right) \right] ~.
\label{sndaction}
\end{equation}
It is convenient to make use of the well-known fact that the three terms of the action are particular cases of Lovelock terms \cite{Lovelock1971}. This allows us to rewrite everything in the language of tensorial forms: the vierbein $e^a$ and spin connection $\omega_{~b}^a$ 1-forms, and the Riemann 2-form,
\begin{equation}
R_{~b}^a = d\omega_{~b}^a + \omega_{~c}^a \wedge \omega_{~b}^c =  \frac{1}{2} R_{~b\mu\nu}^{a}\; dx^{\mu} \wedge dx^{\nu} ~.
\label{Riemann}
\end{equation}
The action (\ref{sndaction}) can be written in terms of these quantities as
\begin{eqnarray}
\mathcal{I} & = & \frac{\lambda\, L^2}{(d-3)! (d-4)} \mathop\int \epsilon_{abcd{f_1}\cdots{f_{d-4}}}\, \left( R^{ab} \wedge R^{cd} + \frac{1}{\lambda\, L^2}\; \frac{d-4}{d-2}\,\; R^{ab} \wedge e^{cd} \right. \nonumber \\ [1em]
& & \left. \qquad\qquad\qquad  + \;\frac{1}{\lambda\, L^4}\,\; \frac{d-4}{d}\,\; e^{abcd} \right) \wedge e^{{f_1}\cdots{f_{d-4}}} ~,
\end{eqnarray}
where $e^{{a_1}\cdots{a_k}} \equiv e^{a_1} \wedge \ldots \wedge e^{a_k}$, and we have used the fact that $\Lambda = - \frac{(d-1)(d-2)}{2\, L^2}$ in $d$ space-time dimensions. If we vary the action with respect to the vierbein,\footnote{The variation of the action with respect to the spin connection is proportional to the torsion $T^a = d e^a + \omega_{~b}^a \wedge e^b$ that we set to zero.} we get
\begin{equation}
\epsilon_{abcd{f_1}\cdots{f_{d-5}}{f_{d-4}}}\, \left( R^{ab} \wedge R^{cd} + \frac{1}{\lambda\, L^2}\; R^{ab} \wedge e^{cd} + \frac{1}{\lambda\, L^4}\; e^{abcd} \right) \wedge e^{{f_1}\cdots{f_{d-5}}} = 0 ~,
\label{GBeom-d}
\end{equation}
which can be written as
\begin{equation}
\epsilon_{abcd{f_1}\cdots{f_{d-5}}{f_{d-4}}}\, \left( R^{ab} - \Lambda_+\; e^{ab} \right) \wedge \left( R^{cd} - \Lambda_-\; e^{cd} \right) \wedge e^{{f_1}\cdots{f_{d-5}}} = 0 ~,
\end{equation}
for two negative values of the effective cosmological constant
\begin{equation}
\Lambda_\pm = - \frac{1 \pm \sqrt{1 - 4\lambda}}{2 \lambda L^2} ~.
\end{equation}
The theory would have a degenerate behaviour whenever the cosmological constants agree. This is accounted for by means of the discriminant
\begin{equation}
\Delta \equiv (\Lambda_+ - \Lambda_-)^2 = \frac{1 - 4\,\lambda}{\lambda^2\, L^4} = 0 \qquad {\rm for} \qquad \lambda = \frac{1}{4} ~.
\label{discr}
\end{equation}
This implies that, for $1 - 4\lambda > 0$, there are two AdS vacua around which we can define our theory. If $1 - 4\lambda < 0$, there is no AdS vacuum. For the exact value $4\lambda = 1$, the theory displays a degenerate behavior due to symmetry enhancement. In the particular case of $d=5$, the symmetry enhances to the full $SO(4,2)$ group and the expression (\ref{sndaction}) gives nothing but the Chern-Simons Lagrangian for the AdS group \cite{Chamseddine1989e} (see also \cite{Zanelli2005}).

It is a well-known fact in Gauss-Bonnet gravity that one of the vacua, the one with the $+$ sign in front of the square root, leads to negative mass black holes with a naked singularity that signals the instability of the vacuum \cite{Boulware1985a}. We are thus led to the remaining branch of solutions. The black hole space-time for this theory reads \cite{Cai2002}
\begin{equation}
ds^2 = - N_{\#}^2\, f(r)\, dt^2 + \frac{dr^2}{f(r)} + \frac{r^2}{L^2}\, d\Sigma_{d-2,k}^2 ~,
\label{bhansatz}
\end{equation}
with $d\Sigma_{d-2,k}^2$ the metric of a $d-2$--dimensional manifold of constant curvature equal to $(d-2) (d-3)\, k$ ($k = 0, \pm 1$ parameterizes the different horizon topologies). For the flat horizon case, $k = 0$, we can use the natural frame,
\begin{eqnarray}
& & e^0 = N_{\#}\, \sqrt{f(r)}\, dt ~, \qquad e^1 = \frac{1}{\sqrt{f(r)}}\, dr ~, \nonumber \\ [0,7em]
& & e^A = \frac{r}{L}\, dx^A ~, \quad {\scriptstyle A = 2, \ldots, d-1} ~,
\label{vierbh}
\end{eqnarray}
and the only non-vanishing components of the spin connection read
\begin{equation}
\omega^0_{~1} = \frac{f'(r)}{2 \sqrt{f(r)}}\; e^0 ~, \qquad \omega^1_{~A}= - \frac{\sqrt{f(r)}}{r}\; e^A ~,
\label{spinbh}
\end{equation}
for torsionless space-time. In turn, the non-vanishing components of the Riemann 2-form, $R^{ab} = d \omega^{ab} + \omega^a_{~c} \wedge \omega^{cb}$, read:
\begin{eqnarray}
& & R^{01} = - \frac{f''(r)}{2} \; e^0 \wedge e^1 ~, \qquad R^{0A} = - \frac{f'(r)}{2 r}\; e^0 \wedge e^A ~, \nonumber \\ [1em]
& & R^{1A} = - \frac{f'(r)}{2 r}\; e^1 \wedge e^A ~, \qquad R^{AB} = - \frac{f(r)}{r^2}\; e^A \wedge e^B ~.
\label{riemannbh}
\end{eqnarray}
It is immediate to see, from these expressions, that an asymptotically AdS solution will satisfy $f(r) \to - \tilde\Lambda\, r^2$ at infinity. Indeed, $\tilde\Lambda$ corresponds to either of the previously discussed $\Lambda_\pm$. If we insert this ansatz into the Euler-Lagrange equations, we get
\begin{equation}
(d-1) \frac{r^4}{L^2} - r^2 (r f'(r) + (d-3) f(r)) + \lambda L^2 f(r) (2 r f'(r) + (d - 5) f(r)) = 0 ~,
\end{equation}
the remaining equations of motion being reducible to the above one. This equation can be solved,
\begin{equation}
f(r) = \frac{r^2}{L^2} \frac{1}{2 \lambda} \left( 1 - \sqrt{1 - 4 \lambda \left( 1 - \frac{r_{+}^{d-1}}{r^{d-1}} \right)} \right) ~,
\label{bhsolution}
\end{equation}
where we have chosen the asymptotic behavior $f(r) \to - \Lambda_-\, r^2$. From this expression we can identify the value for the constant $N_{\#}$,
\begin{equation}
N_{\#}^2 = \left[ \lim_{r \to \infty}\, \frac{L^2}{r^2}\, f(r) \right]^{-1} =\; \frac{1}{2} \left( 1 + \sqrt{1 - 4 \lambda} \right) ~,
\label{Nsost}
\end{equation}
which sets the boundary speed of light to unity. Thermodynamic properties of this black hole solution can be found in \cite{Myers1988,Nojiri2001j,Cvetic2002,Kofinas2006} (and, in the special case $\lambda = 1/4$, in \cite{Banados1994}; see also \cite{Garraffo2008}).

%%%%%%%%%%%%%%%%%%%%%%%%%%%%%%%%%%%%%%%%%%%%%%%%%%%%
\section{Black hole perturbations}
%%%%%%%%%%%%%%%%%%%%%%%%%%%%%%%%%%%%%%%%%%%%%%%%%%%%

In this section we will focus on the 7d case. We will consider perturbations of the metric around the black hole solution (\ref{bhsolution}), along a given direction parallel to the boundary (say, $x^6 \equiv z$) and propagating towards the interior of the geometry. Using the direction $z$ as an axis of symmetry, we can classify the perturbations in helicity representations of the rotation group around it. It is convenient to analyze each case separately.

%%%%%%%%%%%%%%%%%%%%%%%%%%%%%%%%%%%%%%%%%%%%%%%%%%%%
\subsection{Helicity two perturbation}
%%%%%%%%%%%%%%%%%%%%%%%%%%%%%%%%%%%%%%%%%%%%%%%%%%%%

The easiest case is the one with higher helicity. It has been indeed carried out in \cite{Boer2009}, but we report it here for completeness and to explain our quite different framework. For symmetry reasons we can choose the helicity two perturbation, $h_{\mu\nu}(t,r,z)\, dx^\mu dx^\nu$, simply as\footnote{In principle, one should consider $h_{ij}$, with $i < j = 2, \ldots, 6$, but their equations are all decoupled and give rise to the same answer. The remaining helicity two components are $h_{ii} - h_{jj}$, with $i,j = 2, \ldots, 6$, but since diagonal components can be made all equal by rotation, $h_{ii}Ê= 0$.} $h_{23}(t,r,z)\, dx^2 dx^3$. Since we will consider small perturbations, it is convenient to include an infinitesimal parameter $\epsilon$, $h_{23}(t,r,z) = \epsilon\, \phi(t,r,z)$. This can be readily included in the vierbein as $\tilde{e}^a = e^a +\epsilon\,\delta e^a$,
\begin{eqnarray}
& & \tilde{e}\,^0 = N_{\#}\, \sqrt{f}\, dt ~, \qquad\quad \tilde{e}^1 = \frac{1}{\sqrt{f}}\,  dr  ~, \qquad \tilde{e}^K = \frac{r}{L}\,  dx^K ~, \quad {\scriptstyle K = 4, 5, 6} ~, \nonumber \\ [0,7em]
& & \tilde{e}^2 = \frac{r}{L}\,\left( 1 + \frac{\epsilon}{2}\,\phi \right) \left( dx^2 + dx^3 \right)  ~, \qquad \tilde{e}^3 = \frac{r}{L}\,\left( 1 - \frac{\epsilon}{2}\,\phi \right) \left( dx^2 - dx^3 \right) ~,
\label{vierbh-h2}
\end{eqnarray}
From the torsionless condition we can now calculate the first order corrections to the spin-connection, $\tilde{\omega}^a_{~b}=\omega^a_{~b} + \epsilon\,\delta \omega^a_{~b}$, and from them those to the curvature 2-form,
\begin{equation}
\delta R^a_{~b}= d(\delta \omega^a_{~b})+\delta \omega^a_{~c} \wedge \omega^c_{~b} + \omega^a_{~c} \wedge \delta \omega^c_{~b} ~.
\label{deltaRab}
\end{equation}
The first order contribution in $\epsilon$ to the equations of motion can be written as
\begin{eqnarray}
& & \epsilon_{abcdefg}\,  \left( 2 R^{ab} \wedge R^{cd} + \frac{4}{\lambda\, L^2}\; R^{ab} \wedge e^c \wedge e^d + \frac{6}{\lambda\, L^4}\; e^a \wedge e^b \wedge e^c \wedge e^d \right) \wedge e^f \wedge \delta e^g  \nonumber \\ [0.7em]
& & \qquad +\,\epsilon_{abcdefg}\; \delta R^{ab} \wedge \left( 2 R^{cd} + \frac{1}{\lambda\, L^2}\;  e^c \wedge e^d  \right) \wedge e^f \wedge e^g = 0 ~.
\label{EOMh2}
\end{eqnarray}
Consider now the Fourier transform of the perturbation,
\begin{equation}
\phi(t,r,z) = \mathop\int \frac{d\omega}{2\pi}\; \frac{dq}{2\pi}\; \hat\phi(r;k)\; e^{- i \omega t + i q z} ~, \qquad k = (\omega, 0, \ldots, 0, q) ~.
\label{Fourt}
\end{equation}
For the sake of clarity, we omit the $k$ dependence in $\hat\phi$ in what follows.
The equations of motion for $\hat\phi(r)$ result from the insertion of (\ref{vierbh-h2}) and (\ref{deltaRab}) in (\ref{EOMh2}). There is only one independent equation that can be massaged to suit the form
\begin{equation}
K\; \hat\phi'' + K'\; \hat\phi' + K_2\; \hat\phi = 0 ~,
\label{theequation}
\end{equation}
where
\begin{eqnarray}
K & = &  r^3 f(r) \left(-2r^2+\lambda L^2 \left( r f' +2 f\right)\right) ~, \\ [0,5em]
K_2 & = & \frac{ \omega ^2 K}{ N_{\#}^2 f^2 }-\frac{q^2 L^2}{3r^2}  r^3 \left(-6r^2+\lambda L^2 \left(r^2 f''+4rf'+2f\right)\right) ~.
\end{eqnarray}
This is precisely the same result obtained in \cite{Boer2009}, and the discussion follows straightforwardly. Taking $\omega$ and $q$ to infinity, we extract directly the following expression for the speed of the large momentum gravitons on constant $r$ slices
\begin{equation}
c^2_2 = \frac{\omega^2}{q^2} =  \frac{L^2 N_{\#}^2\, f}{3r^2} \frac{\left(-6r^2+\lambda L^2\left(r^2 f'' + 4 r f' + 2 f \right)\right)}{ \left( -2 r^2+\lambda L^2 \left( r f' + 2 f\right)\right)} ~.
\label{c2-h2}
\end{equation}
In \cite{Brigante2008,Brigante2008a} it has been argued that the existence of a maximum in the radial dependence of this speed where it exceeds unity indicates the possibility of bouncing high-momentum gravitons leading to causality violation in the boundary theory. Their arguments extend smoothly to our case. We will rephrase them, for completeness, in what follows. Actually, in the high momentum limit, (\ref{theequation}) can be written as 
\begin{equation}
\tilde g_{eff}^{\mu\nu} \tilde\nabla_\mu \tilde\nabla_\nu \hat\phi = 0 ~,
\end{equation}
where $\tilde\nabla$ is a covariant derivative with respect to the effective geometry given by $\tilde g_{eff}^{\mu\nu} = \Omega^2\; g_{eff}^{\mu\nu}$ with
\begin{equation}
g^{eff}_{\mu\nu}dx^\mu dx^\nu=N_{\#}^2 f(r)\left(-dt+\frac{1}{c_2^2}\,dx^i dx^i\right)+\frac{dr^2}{f(r)} ~,
\label{geff}
\end{equation}
and $\Omega^2$ is a Weyl factor whose particular form is irrelevant for our discussion. In the large momentum limit, a localized wave packet should follow a null geodesic, $x^\mu(s)$, in this effective geometry (\ref{geff}). This follows from standard geometrical optics arguments. If we consider a wave packet with definite momentum
\begin{equation}
\phi=e^{i\Theta(t,r,z)}\phi_{en}(t,r,z) ~,
\end{equation}
where $\Theta$ is a rapidly varying phase and $\phi_{en}$ denotes an `almost constant' envelope, to leading order we find
\begin{equation}
\frac{dx^\mu}{ds}\frac{dx^\nu}{ds}\; g^{eff}_{\mu\nu}=0 ~,
\end{equation}
We have to identify $\frac{dx^\mu}{ds}=g^{\mu\nu}_{eff}\; k_\nu = g^{\mu\nu}_{eff}\,\nabla_\nu\Theta$. As our effective background is symmetric under translations in the $t$ and $z$ directions, we can interpret $\omega$ and $q$ as conserved quantities associated with the corresponding Killing vectors,
\begin{equation}
\omega = k_t = \frac{dt}{ds} N_{\#}^2\, f ~, \qquad q = k_z = \frac{dz}{ds} N_{\#}^2\, \frac{f}{c_2^2} ~.
\label{killing}
\end{equation}
Rescaling the affine parameter as $\tilde{s} = q s/N_{\#}$ (we assume $q\neq0$), we get the following radial equation of motion
\begin{equation}
\left(\frac{dr}{d\tilde{s}}\right)^2=\alpha^2-c_2^2 ~, \qquad \alpha\equiv\frac{\omega}{q} ~.
\end{equation}
This equation describes a particle of energy $\alpha^2$ moving in a potential given by $c_2^2$. If there is a maximum in $c_2^2$, geodesics starting from the boundary can find its way back to the boundary, with turning point $\alpha^2 = c_2^2(r_{\rm turn})$. For a null bouncing geodesic starting and ending at the boundary, we then have
\begin{eqnarray}
\Delta t (\alpha) & = & 2\int_{r_{\rm turn}(\alpha)}^\infty{\frac{\dot{t}}{\dot{r}}\; dr} = \frac{2}{N_{\#}} \int_{r_{\rm turn}( \alpha)}^\infty{\frac{\alpha}{f(r)\sqrt{\alpha^2-c_2^2(r)}}\; dr} ~, \nonumber \\ [0,5em]
\Delta z (\alpha) & = & 2\int_{r_{\rm turn}(\alpha)}^\infty{\frac{\dot{z}}{\dot{r}}\; dr} = \frac{2}{N_{\#}} \int_{r_{\rm turn}( \alpha)}^\infty{\frac{c_2^2}{f(r)\sqrt{\alpha^2-c_2^2(r)}}\; dr} ~,
\label{deltaintegrals}
\end{eqnarray}
where dots indicate derivatives with respect to $\tilde{s}$. Then, as the energy $\alpha$ approaches the value of the speed at the maximum, $\alpha\rightarrow c_{2,{\rm max}}$ ($r_{\rm turn} \rightarrow r_{\rm max}$), the  denominator of the integrand in both expressions diverges and the integrals (\ref{deltaintegrals}) are dominated by contributions from the region near the maximum. Thus, in such a limit we have
\begin{equation}
\frac{\Delta z}{\Delta t} \rightarrow c_{2,{\rm max}} > 1 ~.
\end{equation}
Such geodesics spend a long time near the maximum, traveling with a speed bigger than one. This can be interpreted as bulk disturbances created by local operators in the boundary CFT and we expect microcausality violation in this theory if there exists a bouncing graviton geodesic with $\frac{\Delta t}{\Delta z} > 1$, as in this case. Further discussion on this point can be found in \cite{Brigante2008a,Boer2009}. Actually, it can be shown that the superluminal graviton propagation corresponds to superluminal propagation of metastable quasiparticles in the boundary CFT with $\frac{\Delta z}{\Delta t}$ identified as the group velocity of the quasiparticles.

The existence of such a maximum can be analyzed just by expanding the expression above near the boundary in powers of $1/r$ and picking the sign of the first correction. When this leading correction is positive the maximum exists as when it is negative it does not. In this case we get from (\ref{c2-h2}),
\begin{equation}
c^2_2 \approx 1 - \frac{ \left( 1 + \sqrt{1 - 4 \lambda} - 8 \lambda \right)}{2 (1 - 4 \lambda )}\, \frac{r_+^6}{r^6} + \mathcal{O} \left( \frac{r_+^{12}}{r^{12}} \right) ~,
\end{equation}
thus we see that the perturbation becomes superluminal unless $\lambda$ is bounded from above, as deduced in \cite{Boer2009}, $\lambda \leq \frac{3}{16}$.

%%%%%%%%%%%%%%%%%%%%%%%%%%%%%%%%%%%%%%%%%%%%%%%%%%%%
\subsection{Helicity one perturbation}
%%%%%%%%%%%%%%%%%%%%%%%%%%%%%%%%%%%%%%%%%%%%%%%%%%%%

In order to choose an helicity one perturbation, we can proceed with a gauge fixing ($h_{0a}=0$) and, by symmetry arguments, just turn on the components $h_{12}(t,r,z)$ and $h_{26}(t,r,z)$ (recall that the direction $1$ is the radial one and $6$ is the propagation $z$). Since we will consider small perturbations, we include an infinitesimal parameter $\epsilon$, $h_{12}(t,r,z) = \epsilon\, \phi(t,r,z)$, $h_{26}(t,r,z) = \epsilon\, \psi(t,r,z)$. The helicity one perturbation, thus, can be parameterized as, 
\begin{eqnarray}
& & \tilde{e}\,^0 = N_{\#}\, \sqrt{f}\, dt ~, \qquad\quad \tilde{e}^1 = \frac{1}{\sqrt{f}}\,  dr + \frac{r}{L}\, \epsilon\,\phi\, dx^2 ~, \nonumber \\ [0,7em]
& & \tilde{e}^2 = \frac{r}{L}\, \left( dx^2 + \epsilon\,\psi dz\right) ~, \qquad \tilde{e}^B = \frac{r}{L}\,  dx^B ~, \quad {\scriptstyle B = 3, 4, 5, 6} ~.
\label{vierbh-h1}
\end{eqnarray}
Proceeding as in the previous case we get the following set of algebraic equations 
\begin{eqnarray}
& & \left[-2r^2+\lambda L^2\left(rf'+2f\right)\right]\, \hat\psi = 0 ~, \\ [0.7em]
& & \left[L^2 q^2 N_{\#}^2\, f \left( -2r^2 +\lambda L^2 (r f'+2f)\right) + 2 r^2 w^2 \left( r^2 - 2 \lambda L^2\, f \right)\right]\, \hat\phi = 0 ~,
\end{eqnarray}
where $\hat\psi$ and $\hat\phi$ are radial functions obtained from the Fourier transform of $\psi$ and $\phi$ as in (\ref{Fourt}). The first equation sets the $h_{26}$ component to zero\footnote{This corrects a typo after Eq.(3.30) in \cite{Hofman2009}.}, while the second yields the speed of the gravitons
\begin{equation}
c_1^2 = \frac{\omega^2}{q^2} =  \frac{L^2 N_{\#}^2 f}{2 r^2} \frac{ \left(-2r^2+\lambda L^2 \left(r f' + 2 f\right) \right)}{ \left( -r^2 + 2 \lambda L^2 f\right)} ~.
\end{equation}
Expanding in power series around the boundary,
\begin{equation}
c_1^2 \approx 1 - \frac{\left( 1 + \sqrt{1 - 4 \lambda} + 2 \lambda \right)}{2 (1 - 4 \lambda)}\, \frac{r_+^6}{r^6} + \mathcal{O} \left( \frac{r_+^{12}}{r^{12}} \right) ~.
\end{equation}
Demanding the leading correction to be negative we get $\lambda \geq -2$, in full agreement with the field theory expectation quoted just after (\ref{lambda7d}).

%%%%%%%%%%%%%%%%%%%%%%%%%%%%%%%%%%%%%%%%%%%%%%%%%%%%
\subsection{Helicity zero perturbation}
%%%%%%%%%%%%%%%%%%%%%%%%%%%%%%%%%%%%%%%%%%%%%%%%%%%%

We can proceed in the same way as we did for the other two types of perturbations. The helicity zero perturbation is, anyway, a bit more involved. After gauge fixing, we still need to turn on several components $h_{11} = \psi(t,r,z)$, $h_{22} = h_{33} = h_{44} = h_{55} = \xi(t,r,z)$, $h_{16} = \phi(t,r,z)$, and $h_{66} = \varphi(t,r,z)$ (as before, we will call their Fourier transforms respectively $\hat\psi(r)$, $\hat\xi(r)$, $\hat\phi(r)$ and $\hat\varphi(r)$. The vierbein for this case reads,
\begin{eqnarray}
& & \tilde{e}\,^0 = N_{\#}\, \sqrt{f}\, dt ~, \qquad\quad \tilde{e}^1 = \frac{1}{\sqrt{f}}\, \left( 1 + \frac{\epsilon}{2}\,\psi \right) dr + \frac{r}{L}\, \epsilon\,\phi\, dz ~, \qquad\\
& & \tilde{e}^6 = \frac{r}{L}\, \left( 1 + \frac{\epsilon}{2}\,\varphi \right) dz ~, \qquad \tilde{e}^B = \frac{r}{L}\, \left( 1 + \frac{\epsilon}{2}\,\xi \right) dx^B ~, \quad {\scriptstyle B = 2, 3, 4, 5} ~.
\label{vierbh-h0}
\end{eqnarray}
Keeping only terms quadratic in $\omega$ or $q$ (we will only be interested in the UV behaviour of the theory) the equations of motion for the Fourier components of the perturbations reduce to a merely algebraic set of equations
\begin{eqnarray}
& & (r^2 - 2 \lambda L^2\, f)\; \hat\phi = 0 ~, \qquad (r^2 - 2 \lambda L^2\, f)\; \hat\psi + 2\, (2 r^2 - \lambda L^2\, (2\, f + r f'))\; \hat\xi = 0 ~, \nonumber \\ [0.7em]
& & \left[ 4 \omega^2\, (r^4 - 2 \lambda L^2\, f) + q^2 N_{\#}^2 L^2 (-4 r^2\, f + 2 \lambda L^2\, f\, (2\, f + r\, f')) \right]\; \hat\xi = \nonumber \\ [0.7em]
& & \qquad\qquad\qquad\qquad\qquad\qquad\qquad\qquad\qquad\qquad\qquad\qquad\qquad - \omega^2\, r^2\, (r^2 - 2 \lambda L^2\, f)\; \hat\varphi ~, \nonumber \\ [0.7em]
& & \left[ 5 \omega^2\, (2 r^6 + \lambda L^2 r^2\, (-r^2\, (r\, f' + 6\, f) + 2 \lambda L^2\, f (r\, f' + 2\, f))) + q^2 N_{\#}^2 L^2\, (-10 r^4\, f \right. \nonumber \\ [0.7em]
& & \qquad \left. +\,  \lambda L^2\, f (12 r^3\, f' - r^4\, f'' + 18 r^2\, f - 2 \lambda L^2\, (2 r^2\, f'^2 + f\, (6\, f + 4 r\, f' r^2\, f'')))) \right]\; \hat\xi\nonumber = 0 ~,
\end{eqnarray}
that can be easily solved. From the first three equations we get the components themselves
\begin{eqnarray}
& & \hat\phi = 0 ~, \quad\qquad\quad \hat\psi = - \frac{2 (2 r^2 - \lambda L^2 (2 f + r f'))}{(r^2 - 2 \lambda L^2 f)}\, \hat\xi ~, \nonumber \\ [0.5em]
& & \left[ -2 L^2 r \lambda  f \left( r \left( L^2 \lambda  f''+9 \right) - 4 L^2 \lambda f'\right)+r^2 \left(r^2 \left(L^2 \lambda  f''+10\right)+4 L^4 \lambda ^2 f'^2 \right. \right. \nonumber \\ [0.5em]
& & \left. \left. \qquad -\, 12 L^2 r \lambda  f'\right) +12 L^4 \lambda ^2 f^2 \right] \hat\varphi = 2 L^2 \lambda \left[ r^2 \left(-2 r^2 f''-3 L^2 \lambda  f'^2 \right. \right. \nonumber \\ [0.5em]
& & \left. \left. \qquad\qquad\qquad\qquad\qquad +\,4 r f'\right)+4 r f \left(r \left(L^2 \lambda  f''-1\right) + L^2 \lambda  f'\right)-4 L^2 \lambda  f^2 \right]\,\hat\xi ~, \nonumber
\end{eqnarray}
where we have already substituted the expression for the graviton speed issued from the last equation
\begin{equation}
c_0^2 =  N_{\#}^2\; \frac{F(r)}{5 r^2 (r^2 - 2 L^2 \lambda f) (r (2 r - L^2  \lambda f')-2 L^2 \lambda f)} ~,
\end{equation}
the function $F(r)$ being given by
\begin{eqnarray}
F(r) & = & L^2\, f\, (- 2 L^2 r \lambda\, f\, (r (L^2 \lambda\, f'' + 9) - 4 L^2 \lambda\, f') + r^2 (r^2 (L^2 \lambda\, f'' + 10) \nonumber \\ [0.7em]
& & \qquad + 4\, L^4 \lambda^2\, f'^2 - 12 L^2 r \lambda\, f') + 12 L^4 \lambda^2\, f^2) ~.
\end{eqnarray}
Expanding again in series around the boundary,
\begin{equation}
c_0^2 \approx 1 - \frac{\left(1 + \sqrt{1 - 4 \lambda} + 8 \lambda \right)}{2 (1 - 4 \lambda)}\, \frac{r_+^6}{r^6} + \mathcal{O}\left( \frac{r_+^{12}}{r^{12}} \right) ~,
\label{speed-h0}
\end{equation}
and so the first correction is positive when $\lambda < -\frac{5}{16}$ leading to causality violation.
%%%%%%%%%%%%%
\FIGURE{\includegraphics[width=0.67\textwidth]{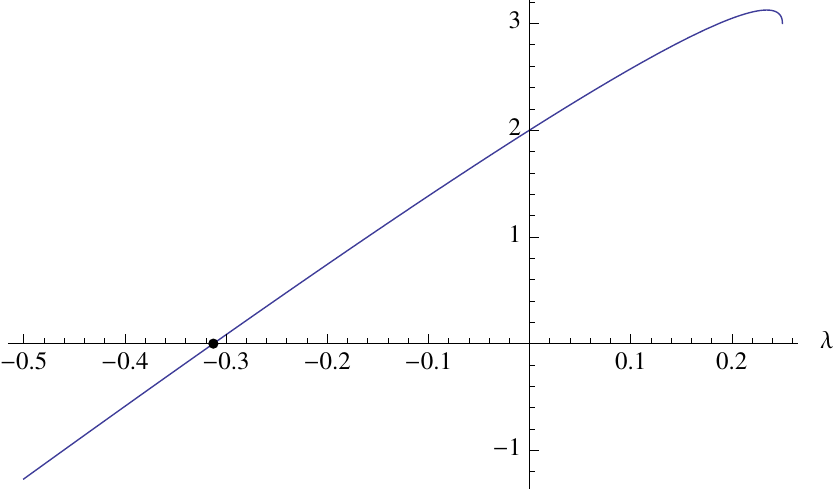}\caption{The numerator of (\ref{speed-h0}) should be positive. This restricts the values of $\lambda$ to $\lambda \geq - 5/16$.}
\label{Helicity0}}
%%%%%%%%%%%%%
\noindent
The lower bound for the GB coupling is then $\lambda \geq -\frac{5}{16}$ as expected from the positivity of the energy analysis carried out in \cite{Boer2009}, see (\ref{lambda7d}). There, it is suggested that a different polarization of the graviton may be responsible for such lower value, and here we confirm that this is the case for the scalar perturbations. The lower bound provided by the helicity one perturbation is less restrictive and, thus, irrelevant.

%%%%%%%%%%%%%%%%%%%%%%%%%%%%%%%%%%%%%%%%%%%%%%%%%%%%
\section{Gravitons colliding shock waves}
%%%%%%%%%%%%%%%%%%%%%%%%%%%%%%%%%%%%%%%%%%%%%%%%%%%%

The previous computations are carried on a black hole background. As such, they are adequate in the context of thermal CFTs. As pointed out in \cite{Hofman2009}, one would expect to be able to perform a similar computation in a zero temperature background. The violation of unitarity driven by a value of the Gauss-Bonnet coupling outside the allowed range, is not an artifact of the finite temperature. An adequate background to perform a computation that is independent of the temperature is given by a pp-wave. In particular, it is easier to consider the simplest case, provided by shock waves \cite{Hofman2009}. They are not subjected to higher derivative corrections \cite{Horowitz1999a}. As such, the AdS shock waves are exact solutions in string theory.

We shall thus consider shock wave backgrounds in Gauss-Bonnet gravity. We will study the scattering of a graviton with a shock wave in AdS. This process is, in a sense, the gravity dual of the energy 1-point function in the CFT \cite{Hofman2008}. We will see that causality violation is again the source of a constraint on the value of $\lambda$. For forbidden values of this coupling, a graviton that is emitted from the boundary comes back and lands outside its light cone. Let us start by considering the case of 7d, which is the setup that we explore in full detail. The results are further generalized for arbitrary higher dimensional space-time in the next section.

%%%%%%%%%%%%%%%%%%%%%%%%%%%%%%%%%%%%%%%%%%%%%%%%%%%%
\subsection{Helicity two graviton}
%%%%%%%%%%%%%%%%%%%%%%%%%%%%%%%%%%%%%%%%%%%%%%%%%%%%

It is more convenient to work in Poincare coordinates, $z = 1/r$. We insist in performing all computations in the formalism used in the previous section since it is significantly simpler than the usual tensorial setup. We define light-cone coordinates\footnote{We then have to change the tangent space metric to $\eta_{00} = \eta_{11} = 0$, $\eta_{01} = \eta_{10} = - \frac12$, $\eta_{AB}=\text{diag}(1,1,\cdots,1), {\scriptstyle A,B = 2, \ldots, 6}$.} $u = t + x^6$ and $v = t - x^6$, and consider a shock wave propagating on AdS along the radial direction,
\begin{equation}
ds^2_{\rm AdS,sw} = ds^2_{\rm AdS} + f(u)\, \varpi(x^a,z)\, du^2 ~, \qquad {\scriptstyle a = 2, 3, 4, 5} ~.
\end{equation}
We should think of $f(u)$ as a distribution with support in $u = 0$, which we will finally identify as a Dirac delta function. As we did in the previous section, we consider an helicity two graviton perturbation, $h_{23}\, dx^2 dx^3$, which we keep infinitesimal $h_{23} = \epsilon\, \phi$,
\begin{equation}
d\tilde{s}^2_{\rm AdS,sw} = \frac{N_{\#}^2}{L^2}\, \frac{-du dv + dx^i dx^i +2 \epsilon \phi \, dx^2dx^3+ L^4 dz^2}{z^2} + f(u)\, \varpi(x^a,z)\, du^2 ~.
\end{equation}
The calculation is very similar to the one in the previous section. We just have to modify the vierbein considered as the input. In this case,
\begin{eqnarray}
& & \tilde{e}\,^0 = \frac{N_{\#}}{L z}\, du ~, \qquad\quad \tilde{e}^1 = \frac{N_{\#}}{L z}\,dv-\frac{L z }{N_{\#}} f(u)\, \varpi(x^a,z)\,du ~, \nonumber \\ [0,5em]
& & \tilde{e}^2 = \frac{N_{\#}}{\sqrt{2} L z} \left(1+\frac{\epsilon}{2}  \phi \right)(dx^2+dx^3) ~, \qquad \tilde{e}^3 = \frac{N_{\#}}{\sqrt{2} L z} \left(1-\frac{\epsilon}{2} \phi \right)(dx^2-dx^3) ~, \nonumber \\ [0,5em]
& & \tilde{e}\,^K =\frac{N_{\#}}{L z}\, dx^K ~, \quad {\scriptstyle K = 4, 5} ~, \qquad \tilde{e}^6 = \frac{N_{\#}L}{z}\, dz ~.
\label{viersw-h2}
\end{eqnarray}
The constant $N_{\#}$ (\ref{Nsost}) is, as we see, related to the radius of the AdS space, and the perturbation depends only on $(u,v,z)$ as before. The shock wave is parameterized by the function $\varpi(x^a,z)$.

Introducing (\ref{viersw-h2}) into the equations of motion for the background ($\epsilon \to 0$) we get an equation for $N_{\#}$ yielding the already known two possible values, and the equation for the shock wave propagating on AdS,
\begin{equation}
8 \varpi - z\, \partial_z \varpi- z^2\, (\partial^2_z \varpi + L^4\; \nabla_\bot^2 \varpi) = 0 ~.
\end{equation}
where $\nabla_\bot^2 = \partial_a \partial^a$ is the Laplacian in the space normal to $z$ and the direction of propagation. There are several possible solutions for this equation. The one we are going to consider is
\begin{equation}
\varpi = \alpha\, N_{\#}^2\, z^4 ~,
\end{equation}
which, as discussed in \cite{Hofman2009}, can be obtained from the black hole background by boosting the solution while keeping its energy constant. The normalization constant $\alpha$ is proportional to the energy density and, as such, must be positive if the solution has a positive mass.

Let us now consider the effective linearized equation of motion for helicity two gravitons in this background. It comes from the $\mathcal{O}(\epsilon)$ contribution to the equations of motion,
\begin{equation}
\frac{5}{z}\, \partial_z \phi - \partial_z^2 \phi + 4 L^4 \left( \partial_u \partial_v  \phi + \alpha f(u)\, L^2 z^6\; \frac{N_{\#}-4\lambda}{N_{\#}-2\lambda}\; \partial^2_v \phi \right) = 0 ~.
\label{swperteq_h2}
\end{equation}
In the large momentum limit and taking the shock wave profile to be a delta function, $f(u) = \delta(u)$, the equation of motion reduces to the usual wave equation $\partial_u \partial_v \phi = 0$ outside $u = 0$. Then, we can consider a wave packet moving with definite momentum on both sides of the shock wave. We can find a matching condition just by integrating over the discontinuity
\begin{equation}
\phi_> = e^{-i P_v\, \alpha\, z^6\; \frac{N_{\#}^2-4\lambda}{N_{\#}^2-2\lambda}}\; \phi_< ~,
\label{matching}
\end{equation}
where we used $P_v = - i \partial_v$. We can find the shift in the momentum in the $z$-direction acting with $P_z = - i \partial_z$,
\begin{equation}
P_z^> = P_z^< - 6 P_v\, \alpha\, z^5\; \frac{N_{\#}^2-4\lambda}{N_{\#}^2-2\lambda} ~.
\end{equation}
If we consider a particle going inside AdS, $P_z > 0$, and if we want it to come back to the boundary after the collision we need
\begin{equation}
P_v\, \alpha\; \frac{N_{\#}^2-4\lambda}{N_{\#}^2-2\lambda} > 0 ~.
\end{equation}
But we know that $\alpha > 0$ (since the black hole has positive mass) and $P_v = - \frac12 P^u < 0$ (since $P^u = P^0 + P^6$ must be positive for the energy to be so); then we need  
%%%%%%%%%%%%%
\FIGURE{\includegraphics[width=0.49\textwidth]{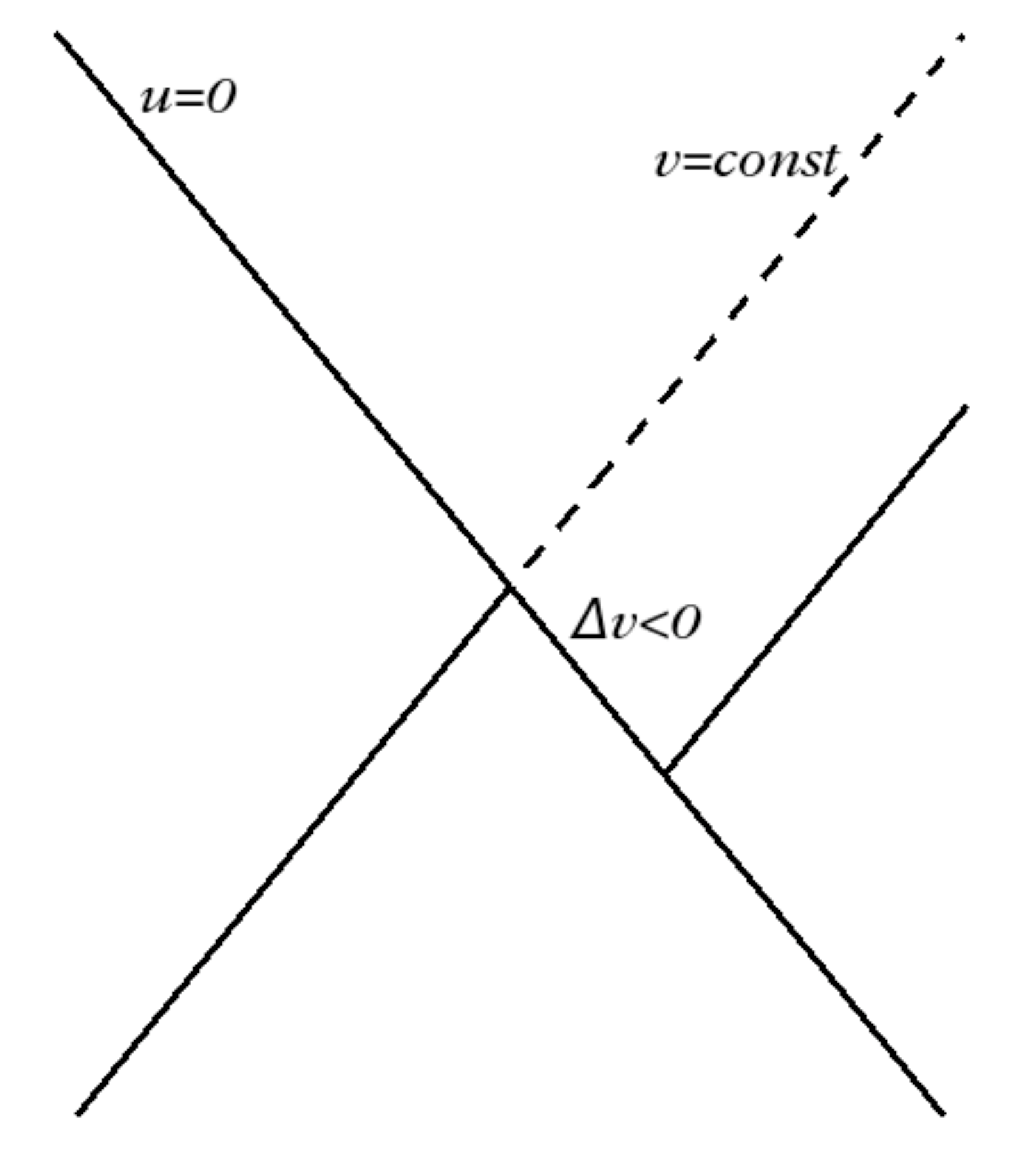}\caption{The line $u=0$ corresponds to the shock wave while the line $v=const.$ corresponds to the graviton. After the collision, if $\Delta v<0$, the particle lands outside its light-cone.}\label{shockwave}}
%%%%%%%%%%%%%
%
\begin{equation}
N_{\#}^2-4\lambda< 0 ~.
\end{equation}
since the denominator is positive for $\lambda < \frac14$, $N_{\#}^2 - 2 \lambda = N_{\#}^2\sqrt{1-4\lambda} > 0$. The numerator changes sign at $\lambda = \frac{3}{16}$. When $\lambda$ is bigger than this critical value the graviton can make its way back to the boundary and, as we can read from (\ref{matching}), it comes back shifted in the $v$-direction a negative amount (see Figure \ref{shockwave})
\begin{equation}
\Delta v = \alpha\, z^6\, \frac{N_{\#}^2 - 4 \lambda}{N_{\#}^2 - 2 \lambda} < 0 ~.
\end{equation}
The graviton lands, at the boundary, outside its own light-cone. This is an explicit break up of causality. We conclude that the theory violates causality unless the Gauss-Bonnet coupling $\lambda$ is bounded from above, $\lambda \leq \frac{3}{16}$.

%%%%%%%%%%%%%%%%%%%%%%%%%%%%%%%%%%%%%%%%%%%%%%%%%%%%
\subsection{Helicity one graviton}
%%%%%%%%%%%%%%%%%%%%%%%%%%%%%%%%%%%%%%%%%%%%%%%%%%%%

We can proceed in an almost identical manner with the other two possible polarizations of the graviton. The only difference is that we have to solve first the equations for several components of the perturbations, classified by the little group about the propagation direction, but at the end of the day the system always reduce to just one equation of the type of (\ref{swperteq_h2}). In the helicity one case we considered the following frame
\begin{eqnarray}
& & \tilde{e}\,^0 = \frac{N_{\#}}{L z}\, du ~, \qquad\quad \tilde{e}^1 = \frac{N_{\#}}{L z}\,dv-\frac{L z }{N_{\#}} f(u)\, \varpi(x^a,z)\,du ~, \nonumber \\ [0,5em]
& & \tilde{e}^2 = \frac{N_{\#}}{L z}\left(dx^2+\epsilon\, \psi \frac12\left(du-dv\right)\right) ~,  \\ [0,5em]
& & \tilde{e}\,^B =\frac{N_{\#}}{L z}\, dx^B ~, \quad {\scriptstyle B = 3, 4, 5} ~, \qquad \tilde{e}^6 = \frac{N_{\#}}{z}\,\left( L dz + \epsilon\, \phi\, \frac{dx^2}{L}\right)\nonumber ~.
\label{viersw-h1}
\end{eqnarray}
Considering the large momentum limit, in this case, the component $\psi$ is set to zero by the equations of motion while we get the following equation for $\phi$:
\begin{equation}
\partial_u \partial_v \phi + \alpha\, f(u)\, L^2 z^6\; \frac{N_{\#}^2+\lambda}{N_{\#}^2-2\lambda}\; \partial^2_v\phi = 0 .
\label{swperteq_h1}
\end{equation}
Therefore, following the same reasoning as in the previous subsection, we find that there is causality violation whenever $N_{\#}^2+\lambda<0$, that is, $\lambda < -2$. 

%%%%%%%%%%%%%%%%%%%%%%%%%%%%%%%%%%%%%%%%%%%%%%%%%%%%
\subsection{Helicity zero graviton}
%%%%%%%%%%%%%%%%%%%%%%%%%%%%%%%%%%%%%%%%%%%%%%%%%%%%

In the helicity zero case, the symmetry is less restricting and the basis of one forms considered is
\begin{eqnarray}
& & \tilde{e}\,^0 = \frac{N_{\#}}{L z}\, \left( \left( 1 + \frac{\epsilon}{4}\, \varphi \right) du - \frac{\epsilon}{4} \varphi\, dv \right) ~, \qquad \tilde{e}\,^B =\frac{N_{\#}}{L z}\,\left(1+\frac{\epsilon}{2}\, \xi\right) dx^B ~, \quad {\scriptstyle B = 2, 3, 4, 5} ~, \nonumber \\ [0,5em]
& & \tilde{e}^1 = \frac{N_{\#}}{L z}\, \left(\left(1+\frac{\epsilon}{4}\varphi\right)dv-\frac{\epsilon}{4}\, \varphi\, du\right)-\frac{L z}{N_{\#}} f(u)\, \varpi(x^a,z)\,\left(\left(1-\frac{\epsilon}{4}\varphi\right) du+\frac{\epsilon}{4} \varphi dv\right) ~, \nonumber \\ [0,5em]
& &  \tilde{e}^6 = \frac{N_{\#}L}{z}\,\left(1+\frac{\epsilon}{2}\, \xi\right) dz+\frac{N_{\#}}{L z}\epsilon\, \phi \frac12(du-dv) ~.
\label{viersw-h0}
\end{eqnarray}
In the large momentum limit, the equations of motion are easily solved by setting $\phi=0$ and $\psi = - 4 \xi$. The remaining equations involving $\varphi$ and $\xi$ read:
\begin{eqnarray}
4 L^2 z^6 \alpha\, f(u) \left( N_{\#}^2 + 16 \lambda \right) \partial_v^2 \xi - \left( N_{\#}^2 - 2 \lambda \right) \left( -4\, \partial_u\, \partial_v\, \xi + \partial_v^2 \varphi + 2 \partial_u \partial_v \varphi + \partial^2_u \varphi \right) = 0 ~, \nonumber \\ [0,7em]
16 L^2 z^6 \alpha\, f(u) \left( N_{\#}^2 + \lambda \right) \partial_v^2 \xi + \left( N_{\#}^2 - 2 \lambda \right) \left( 16\, \partial_u\, \partial_v \xi + \partial_v^2 \varphi + 2 \partial_u \partial_v\, \varphi + \partial_u^2 \varphi \right) = 0 ~, \quad \nonumber
\end{eqnarray} 
and adding them we get an equation for $\xi$ of the type studied before
\begin{equation}
\partial_u \partial_v \phi + \alpha\, f(u)\, L^2 z^6\; \frac{N_{\#}^2+4\lambda}{N_{\#}^2-2\lambda}\partial^2_v\phi = 0 ~,
\label{swperteq_h0}
\end{equation}
where causality breaking up is seen to arise when $\lambda < - \frac{5}{16}$. All the results match perfectly with those coming from the black hole calculation. 

%%%%%%%%%%%%%%%%%%%%%%%%%%%%%%%%%%%%%%%%%%%%%%%%%%%%
\section{Generalization to higher dimensions}
%%%%%%%%%%%%%%%%%%%%%%%%%%%%%%%%%%%%%%%%%%%%%%%%%%%%

There are at least two arguments to prevent ourself to continue climbing to upper dimensionalities. Both the critical dimension of string theory and the absence of an holographic renormalization result for the central charges of the conjectural higher dimensional CFTs. However, it is tempting to explore some of the results that we can already deal with. If AdS/CFT is valid for arbitrary dimensions, even if it isn't the case, it is still meaningful to explore causality constraints on perturbations corresponding to different helicities in black holes in AdS with flat horizons. On the CFT side, if these higher dimensional CFTs exist, a conformal collider setup would lead us to (\ref{t2andt4-d}), and positive energy conditions shall immediately arise. Let us analyze these issues further in the framework of Gauss-Bonnet theory.

%%%%%%%%%%%%%%%%%%%%%%%%%%%%%%%%%%%%%%%%%%%%%%%%%%%%%%%%%%%%%%%
\subsection{Black hole perturbations}
%%%%%%%%%%%%%%%%%%%%%%%%%%%%%%%%%%%%%%%%%%%%%%%%%%%%%%%%%%%%%%%

We shall start from the equations of motion of Gauus-Bonnet theory in $d$ space-time dimensions (\ref{GBeom-d}), which we rewrite as
\begin{equation}
\mathcal{E}^{(d)}_a = \epsilon_{a{f_1}\cdots{f_{d-1}}}\, \sum_{n=0}^2 c_n\, R^{f_1\cdots f_{2n}} \wedge e^{{f_{2n+1}}\cdots{f_{d-1}}} = 0 ~,
\label{GBeom-d-sum}
\end{equation}
where $R^{f_1\cdots f_{2n}} \equiv R^{f_1f_2} \wedge \cdots \wedge R^{f_{2n-1}f_{2n}}$, and the coefficients are $c_0 = 1/\lambda L^4$, $c_1 = 1/\lambda L^2$ and $c_2 = 1$. The linear order contribution coming from small perturbations read
\begin{eqnarray}
\delta \mathcal{E}^{(d)}_a & = & \epsilon_{a{f_1}\cdots{f_{d-1}}}\, \sum_{n=0}^2 c_n\, \left[ n\, \delta R^{f_1f_2} \wedge R^{f_3\cdots f_{2n}} \wedge e^{{f_{2n+1}}\cdots{f_{d-1}}} \right. \nonumber \\ [0,5em]
& & \left. \quad\qquad\quad +\, (d-2n-1)\, R^{f_1\cdots f_{2n}} \wedge e^{{f_{2n+1}}\cdots{f_{d-2}}} \wedge \delta e^{f_{d-1}} \right] = 0 ~.
\label{lineareq}
\end{eqnarray}
The first thing we have to realize in order to carry out this calculation is that the relevant contributions to order $\omega^2$, $q^2$ and $\omega\, q$, come from derivatives along the directions $e^0$ and $e^{d-1}$ (recall that $x^{d-1}\equiv z$ is the direction of propagation we chose for the perturbation). In the simplest case, for helicity two perturbations (\ref{vierbh-h2}), these contributions have only an effect on the expressions of $\delta \omega^{02}$, $\delta \omega^{03}$, $\delta \omega^{(d-1)2}$ and $\delta \omega^{(d-1)3}$. Since we are at the linearized level, we conclude that the only non-trivial contributions to order $\omega^2$, $q^2$, $\omega\, q$ come from their exterior derivative (the second term in (\ref{lineareq}) is also irrelevant),
\begin{eqnarray}
& & \delta R^{02} \approx d (\delta \omega^{02}) = - \frac{\omega^2}{2 N_{\#}^2 f}\;\phi\; e^0 \wedge e^2 + \frac{\omega\, q L}{2 r N_{\#} \sqrt{f}}\; \phi\; e^{d-1}\wedge e^2 ~, \label{spcon-1} \\ [0,5em]
& & \delta R^{(d-1)2} \approx d (\delta \omega^{(d-1)2}) = - \frac{q^2 L^2}{2 r^2}\; \phi\; e^{d-1} \wedge e^2 - \frac{\omega\, q L}{2 r N_{\#} \sqrt{f}}\; \phi\; e^0\wedge e^2 ~,
\label{spcon-2}
\end{eqnarray}
where the symbol $\approx$ only pays attention to those contributions relevant to compute the propagation speed of a boundary perturbation. There are analog expressions (with the opposite sign) for the components with a leg along the $e^3$ direction. Notice that the cosmological constant term is irrelevant for our purposes,
\begin{equation}
\delta \mathcal{E}^{(d)}_a \approx \epsilon_{a{f_1}\cdots{f_{d-1}}}\, d (\delta \omega^{f_1f_2}) \wedge \left( 2 R^{f_3f_4} + \frac{1}{\lambda L^2} e^{f_3f_4} \right) \wedge e^{{f_5}\cdots{f_{d-1}}} ~.
\label{linnear}
\end{equation}
Recalling from (\ref{riemannbh}) that the curvature 2-form of the black-hole space-time is proportional to $e^a \wedge e^b$, it is immediate to see that the only non-vanishing contributions are those with $d (\delta \omega^{ab})$ behaving equally,
\begin{eqnarray}
& & d (\delta \omega^{02}) \approx - \frac{\omega^2}{2 N_{\#}^2 f}\;\phi\; e^0 \wedge e^2 ~, \quad\qquad\quad d (\delta \omega^{03}) \approx \frac{\omega^2}{2 N_{\#}^2 f}\;\phi\; e^0 \wedge e^3 ~, \nonumber \\ [0,5em]
& & d (\delta \omega^{(d-1)2}) \approx - \frac{q^2 L^2}{2 r^2}\; \phi\; e^{d-1} \wedge e^2 ~, \qquad d (\delta \omega^{(d-1)3}) \approx \frac{q^2 L^2}{2 r^2}\; \phi\; e^{d-1} \wedge e^3 ~, \nonumber
\end{eqnarray}
Some of the equations of motion are satisfied trivially. Indeed, notice that $\delta \mathcal{E}^{(d)}_a \approx 0$ trivially, unless  $a = 2$ or $3$. This is due to cancellations of contributions coming from $d(\delta \omega^{02})$ and $d(\delta \omega^{03})$ (similarly, $d(\delta \omega^{(d-1)2})$ and $d(\delta \omega^{(d-1)3})$). Moreover, by symmetry, the two non-trivial equations just differ by a global sign. We must then focus on a single component of (\ref{linnear}), say, $\delta \mathcal{E}^{(d)}_3 = 0$,
\begin{eqnarray}
\delta \mathcal{E}^{(d)}_3 & \approx & 2\, \epsilon_{{023f_3}\cdots{f_{d-1}}}\, d(\delta \omega^{02}) \wedge \left( 2 R^{f_3f_4} + \frac{1}{\lambda L^2} e^{f_3f_4} \right) \wedge e^{{f_5}\cdots{f_{d-1}}}  \nonumber \\ [0,5em]
& & \qquad +\, 2\, \epsilon_{{(d-1)23f_3}\cdots{f_{d-1}}}\, d(\delta \omega^{(d-1)2}) \wedge \left( 2 R^{f_3f_4} + \frac{1}{\lambda L^2} e^{f_3f_4} \right) \wedge e^{{f_5}\cdots{f_{d-1}}} ~. \nonumber
\end{eqnarray}
To proceed, the only thing to worry about is where are the $0$ and $1$ indices, since depending on them the curvature 2-form components change (\ref{riemannbh}). Notice that the first (second) line gives the $\omega^2$ ($q^2$) contribution. The former reads
\begin{eqnarray}
\delta \mathcal{E}^{(d)}_{3,\omega} & \approx & - \frac{\omega^2}{N_{\#}^2 f}\; \phi\; \epsilon_{{0123f_4}\cdots{f_{d-1}}}\, e^0 \wedge e^2 \wedge \left[ 2 \left( 2 R^{1f_4} + \frac{1}{\lambda L^2} e^{1f_4} \right) \wedge e^{f_5} \right. \nonumber \\ [0,5em]
& & \left. \quad\qquad\quad +\, (d - 5)\; e^1 \wedge \left( 2 R^{f_4f_5} + \frac{1}{\lambda L^2} e^{f_4f_5} \right) \right] \wedge e^{{f_6}\cdots{f_{d-1}}} \nonumber \\ [0,5em]
& = & - (d-4)!\; \left[ \frac{2 f'}{r} + \frac{2 (d - 5) f}{r^2} - \frac{d-3}{\lambda L^2}  \right] \frac{\omega^2}{N_{\#}^2 f}\; \phi ~, \nonumber
\end{eqnarray}
while the latter
\begin{eqnarray}
\delta \mathcal{E}^{(d)}_{3,q} & \approx & \frac{q^2 L^2}{r^2}\; \phi\; \epsilon_{{(d-1)0123f_5}\cdots{f_{d-1}}}\, e^{d-1} \wedge e^2 \wedge \left[ 2 \left( 2 R^{10} + \frac{1}{\lambda L^2} e^{10} \right) \wedge e^{f_5f_6} \right. \nonumber \\ [0,5em]
& & - 2 (d-5) \left( 2 R^{1f_5} + \frac{1}{\lambda L^2} e^{1f_5} \right) \wedge e^{0f_6}  + 2 (d-5) \left( 2 R^{0f_5} + \frac{1}{\lambda L^2} e^{0f_5} \right) \wedge e^{1f_6}  \nonumber \\ [0,5em]
& & \left. + (d-5) (d-6) \left( 2 R^{f_5f_6} + \frac{1}{\lambda L^2} e^{f_5f_6} \right) \wedge e^1 \wedge e^0  \right] \wedge e^{{f_7}\cdots{f_{d-1}}} \nonumber \\ [0,5em]
& = & (d-5)!\; \left[ 2 f'' + \frac{4 (d-5) f'}{r} + \frac{2 (d-5) (d-6) f}{r^2} - \frac{(d-3) (d-4)}{\lambda L^2} \right] \frac{q^2 L^2}{r^2}\; \phi ~, \nonumber
\end{eqnarray}
where in the last step of both expressions we got rid of a $d-1$ volume form, to ease the notation. It is convenient to define the following functionals, $\mathcal{C}^{(k)}_d[f,r]$, involving up to $k$th-order derivatives of $f$:
\begin{eqnarray}
\mathcal{C}^{(0)}_d[f,r] & = & \lambda L^2 \sum_{n=1}^2 n\, c_n \frac{(-1)^n}{r^{2n}}\, f^{n-1} = \frac{1}{r^4} \left[ -r^2 + 2 \lambda L^2\, f \right] ~, \\ [0,3em]
\mathcal{C}^{(1)}_d[f,r] & = & \lambda L^2 \sum_{n=1}^2 n\,c_n \frac{(-1)^n}{r^{2n}}\; \left( r\, (f^{n-1})' + (d-2n-1)\, f^{n-1} \right) \nonumber \\ [0,3em]
& = & \frac{1}{r^4} \left[ - (d-3) r^2 + 2 \lambda L^2 (r\, f' + (d-5)\, f) \right] ~, \\ [0,3em]
\mathcal{C}^{(2)}_d[f,r] & = & \lambda L^2\sum_{n=1}^2 c_n n \frac{(-1)^n}{r^{2n}}\, \left( r^2\, (f^{n-1})'' + 2 (d-2n-1) r\, (f^{n-1})' \right. \nonumber \\ [0,3em]
& & \qquad \qquad \qquad \left. +\; (d-2n-1) (d-2n-2)\, f^{n-1} \right) \nonumber \\ [0,3em]
& = & \frac{1}{r^4} \left[ - (d-3) (d-4) r^2 + 2 \lambda L^2 (r^2\, f'' + (d-5) (2 r\, f' + (d-6)\,f)) \right] ~.
\end{eqnarray}
The usefulness of these expressions is manifest in the compact form of the speed of the helicity two graviton that can be simply written as
\begin{equation}
c_2^2 = \frac{N_{\#}^2 L^2 f}{(d-4)\, r^2}\; \frac{\mathcal{C}^{(2)}_d[f,r]}{\mathcal{C}^{(1)}_d[f,r]} ~,
\end{equation}
and expanding around the boundary 
\begin{equation}
c_2^2 = 1+\frac{\left(1+\sqrt{1-4 \lambda }\right) \left(2(d-1)-(d^2-5d+10) \sqrt{1-4 \lambda} \right)}{2 (d-4) (d-3) (1-4 \lambda )}\frac{r_+^{d-1}}{r^{d-1}}+\mathcal{O}\left(\frac{r_+^{2d-2}}{r^{2d-2}}\right)
\end{equation}
that we can express in a more suitable way for latter purposes:\footnote{We thank Diego Hofman for suggesting this possibility to us.}
\begin{equation}
c_2^2 = 1 - \frac{1}{(1-4\lambda)} \left( N_{\#}^2 - \frac{2(d^2-5d+10)}{(d-3)(d-4)}\, \lambda \right) \frac{r_+^{d-1}}{r^{d-1}} + \mathcal{O}\left(\frac{r_+^{2d-2}}{r^{2d-2}}\right) ~.
\end{equation}
Requiring this leading correction to the speed of the helicity two graviton to be negative, we find the bound on the Gauss-Bonnet parameter
\begin{equation}
\lambda \leq \frac{(d-4) (d-3) \left( d^2 - 3 d + 8 \right)}{4 \left(d^2 - 5 d + 10 \right)^2} ~.
\label{upbound-d}
\end{equation}
It is immediate to verify that the lower dimensional values coincide with previous results in the literature; respectively $\lambda \leq 9/100$ for $d = 5$ \cite{Brigante2008,Brigante2008a}, and $\lambda \leq 3/16$ for $d = 7$ \cite{Boer2009}. Indeed, this upper bound coincides with the general $d$ expression found in \cite{Ge2009}

It is a bit more involved to work out the other helicities but the same procedure can be followed. For the helicity one case we have to solve the equations of motion for two components, but one of them always vanishes. The remaining equation yields directly the speed of the graviton as in the previous case
\begin{equation}
c_1^2 = \frac{N_{\#}^2 L^2 f}{(d-3)\,r^2}\; \frac{\mathcal{C}^{(1)}_d[f,r]}{\mathcal{C}^{(0)}_d[f,r]} ~.
\end{equation}
Expanding again close to the boundary
\begin{equation}
c_1^2 = 1 + \frac{\left(1+\sqrt{1-4 \lambda }\right) \left(d -1 -2\sqrt{1-4\lambda}\right)}{2 (d-3) (1-4 \lambda )}\; \frac{r_+^{d-1}}{r^{d-1}}+\mathcal{O}\left(\frac{r_+^{2d-2}}{r^{2d-2}}\right) ~.
\end{equation}
that can be nicely rewritten as
\begin{equation}
c_1^2=1-\frac{1}{(1-4\lambda)}\left(N_{\#}^2+\frac{4}{(d-3)}\lambda\right)\; \frac{r_+^{d-1}}{r^{d-1}}+\mathcal{O}\left(\frac{r_+^{2d-2}}{r^{2d-2}}\right) ~,
\end{equation}
giving a lower bound for the Gauss-Bonnet coupling
\begin{equation}
\lambda \geq -\frac{1}{16} (d-3) (d+1) ~,
\end{equation}
provided we demand causal propagation in the boundary. This result has not being found earlier and, again, it coincides with the 5d value, $\lambda \geq - 3/4$, obtained in \cite{Buchel2009a}.

The more involved case is the helicity zero one, where we have to find the four components of the perturbations. One of them ($\phi$) is set to zero by the equations of motion, while other two can be written in terms of only one degree of freedom, as it happened in the 7d case analyzed above,
\begin{equation}
\psi = -\frac{\mathcal{C}^{(1)}_d[f,r]}{\mathcal{C}^{(0)}_d[f,r]}\;\xi ~, \qquad 
\varphi = \left(\frac{q^2}{\omega^2}\frac{ N_{\#}^2 L^2 f(r)}{r^2}\; \frac{\mathcal{C}^{(1)}_d[f,r]}{\mathcal{C}^{(0)}_d[f,r]}-(d-3)\right)\,\xi ~.
\end{equation}
When we substitute these expressions into the equations of motion, only one of them remains linearly independent and gives the speed of the helicity zero graviton
\begin{equation}
c_0^2 = \frac{N_{\#}^2 L^2 f}{(d-2)\,r^2}\left( \frac{2\,\mathcal{C}^{(1)}_d[f,r]}{\mathcal{C}^{(0)}_d[f,r]} - \frac{\mathcal{C}^{(2)}_d[f,r]}{\mathcal{C}^{(1)}_d[f,r]}\right) ~.
\end{equation}
From this expression we get the more stringent lower bound for $\lambda$, since the condition we obtain is always more restrictive than that of the helicity one for any dimension. Expanding one more time about the boundary and keeping just the first correction to unity
\begin{equation}
c_0^2 = 1-\frac{\left(1+\sqrt{1-4 \lambda }\right) \left(d \left(2- \sqrt{1-4 \lambda }\right) -2- \sqrt{1-4 \lambda }\right)}{2 (d-3) (1-4 \lambda)}\; \frac{r_+^{d-1}}{r^{d-1}}+\mathcal{O}\left(\frac{r_+^{2d-2}}{r^{2d-2}}\right) ~,
\end{equation}
that can be, again, rewritten in a very helpful form (see below)
\begin{equation}
c_0^2 = 1-\frac{1}{(1-4\lambda)}\left(N_{\#}^2+\frac{2 (d+1)}{(d-3)}\lambda\right)\frac{r_+^{d-1}}{r^{d-1}}+\mathcal{O}\left(\frac{r_+^{2d-2}}{r^{2d-2}}\right) ~,
\end{equation}
this leading to the result:
\begin{equation}
\lambda \geq - \frac{(d-3) (3 d-1)}{4 (d+1)^2}
\end{equation}
The arguments showing that the lower and upper bound of $\lambda$ come from, respectively, helicity zero and helicity two perturbations or, conversely, positivity of the energy in the dual CFT, would lead us to a formula valid for any dimension $d \geq 5$ (below 5d, the Gauss-Bonnet term either is a total derivative or it identically vanishes),
\begin{equation}
-\frac{(d-3) (3 d-1)}{4 (d+1)^2} \leq \lambda \leq \frac{(d-4) (d-3) \left(d^2-3 d+8\right)}{4 \left(d^2-5 d+10\right)^2} ~.
\label{uplambd}
\end{equation}
%
%%%%%%%%%%%%%
\FIGURE{\includegraphics[width=0.67\textwidth]{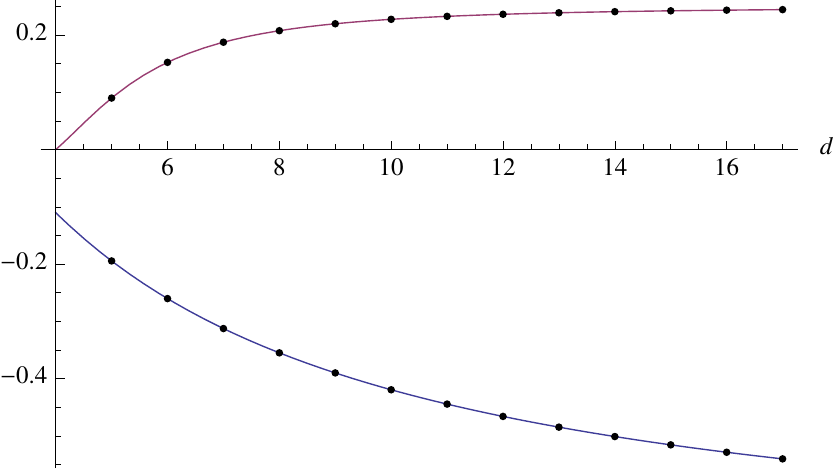}\caption{Upper and lower bound values for $\lambda$. The upper curve corresponds to helicity two modes, while the lower one is due to helicity zero perturbations. The window monotonically increases an asymptotically approaches its maximum range $-3/4 \leq \lambda \leq 1/4$ for infinite dimensional theories.}
\label{bounds}}
%%%%%%%%%%%%%
\noindent
This formula reproduces earlier results \cite{Brigante2008,Brigante2008a,Boer2009,Ge2009} for the upper bound, and generalizes the lower bound to arbitrary dimensions.

There are several comments we would like to make about this result. Besides its remarkable simplicity and smoothness, we see that $\lambda_{\rm max}$ asymptotically approaches $1/4$, when $d \to \infty$. This may be expected. One can show that $\lambda_{\rm max}$ is a monotonically increasing function, but there is an obstruction precisely at $\lambda = 1/4$, as we discussed around (\ref{discr}). It is more striking what happens to the lower bound. There is no critical negative value of $\lambda$, at least manifestly. Thus, naively one might expect that $\lambda_{\rm min} \to - \infty$ in the infinite dimensional limit. However, we obtain $\lambda_{\rm min} \to - 3/4$ (see Figure \ref{bounds}). We think that this asymptotic behavior calls for a deeper understanding.

One of the main consequences of a positive $\lambda$ is the violation of the so-called KSS bound for the shear viscosity to entropy density ratio \cite{Kovtun2005}. As pointed out in \cite{Brigante2008},
\begin{equation}
\frac{\eta}{s} = \frac{1}{4\pi} \left( 1 - 2 \frac{d-1}{d-3}\, \lambda \right) ~,
\end{equation}
for a CFT plasma dual to a Gauss-Bonnet theory. We see that the maximal violation of the KSS bound happens for conjectural 8d CFTs, the minimum value of $\eta/s$ asymptotically approaching the ratio $\eta/s = 1/8\pi$. Whether there exist higher dimensional CFTs with a finite temperature regime admitting a hydrodynamical description with such low values of $\eta/s$ is, of course, an open problem. A warning remark is however worth at this point. The low energy effective gravity action used in these computations is strictly valid in the region of large central charges when their relative differences are very small. Thus, finite values of the GB coupling, $\lambda \sim 1$, are not fully reliable.
\vskip3mm

%%%%%%%%%%%%%
\FIGURE{\includegraphics[width=0.67\textwidth]{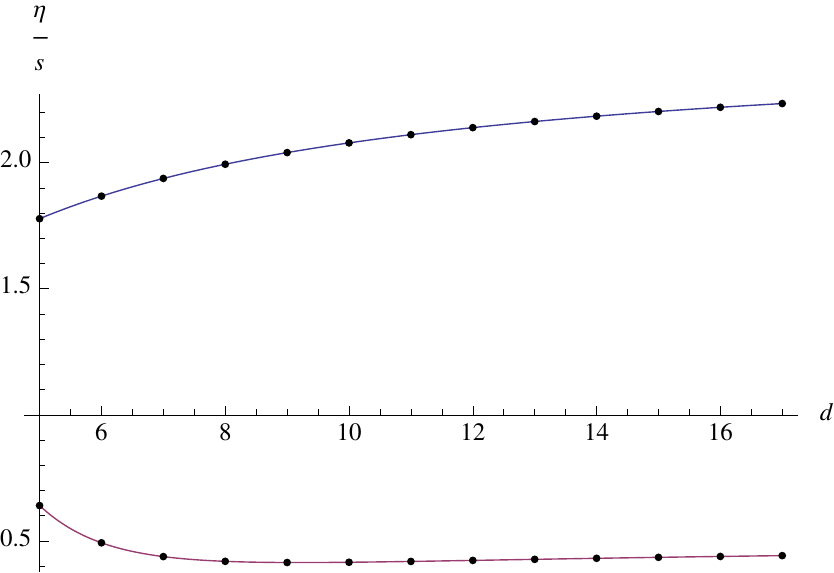}\caption{Upper and lower values for $\eta/s$. The upper curve corresponds to helicity zero, while the lower one is due to helicity two perturbations. The window asymptotically approaches its maximum range $1/2 \leq 4\pi\,\eta/s \leq 5/2$ for infinite dimensions.}
\label{eta-s-vs-d}}
%%%%%%%%%%%%%
The very existence of a negative lower bound for $\lambda$ seems to imply, naively, that there is an upper bound for $\eta/s$ in strongly coupled CFTs (see Figure \ref{eta-s-vs-d}). It is well-known that $\eta/s \to \infty$ is a generic feature in weakly coupled theories but, to the best of our knowledge, there is no a priori reason that tells us why the strongly coupled value should be $1/4\pi$ or differ by a factor of order 1. This seems to be a possible interpretation of our result: no matter the dimensionality of a CFT, its strongly coupled plasma will have a very small shear viscosity to entropy density ratio. To put this conclusion in more firm grounds, however, one should study more carefully the effect of higher curvature corrections and have a deeper understanding on the nature of higher dimensional CFTs. If AdS/CFT is valid for arbitrary dimensions, equation (\ref{uplambd}) gives a qualitative prediction for the bound on the difference of central charges (the prediction becoming quantitative once the holographic renormalization calculus is performed).

%%%%%%%%%%%%%%%%%%%%%%%%%%%%%%%%%%%%%%%%%%%%%%%%%%%%%%%%%%%%%%%
\subsection{Shock waves}
%%%%%%%%%%%%%%%%%%%%%%%%%%%%%%%%%%%%%%%%%%%%%%%%%%%%%%%%%%%%%%%

The shock wave calculation can also be generalized to general $d$ space-time dimensions. The relevant shock wave solution is $\varpi = \alpha\, N_{\#}^2\, z^{d-3}$. The proceedure is almost the same as in the previous section, just a bit more complicated since the symmetry of the background is lower than in the black hole solution. As before, since we are only interested in the high momentum limit, we keep only contributions of the sort $\partial^2_v\phi$, $\partial_u\partial_v\phi$ and $\partial_u^2\phi$. These contributions come again from the exterior derivative of the perturbation of the spin connection. In the helicity two case
\begin{eqnarray}
& & d(\delta \omega^{02}) = \frac{L^2\, z^2}{N_{\#}^2}\, \left[ \partial_v^2 \phi\; e^1 \wedge e^2 + \left( \partial_u \partial_v \phi + \alpha f(u) L^2 z^{d-1}\, \partial_v^2 \phi \right)\, e^0 \wedge e^2 \right] ~, \label{dw02} \\ [0,7em]
& & d(\delta \omega^{12}) = \frac{L^2\, z^2}{N_{\#}^2}\, \left[ \left( \partial_u \partial_v \phi + \alpha f(u) L^2 z^{d-1}\, \partial_v^2\phi \right)\, e^1\wedge e^2 + \left( \cdots \right)\, e^0\wedge e^2 \right] ~, \label{dw12}
\end{eqnarray}
the ellipsis being used in the second expression since the corresponding term does not contribute to the equations of motion. The components with index 3 instead of 2 are the only remaining ones non-vanishing, and they are obtained changing $\phi\rightarrow- \phi$. The other thing we need is the curvature 2-form of the background metric, that can be written as
\begin{equation}
R^{ab} = \Lambda (e^a\wedge e^b + X^{ab}) ~,
\label{RabXab}
\end{equation}
where $\Lambda=-\frac{1}{L^2 N_{\#}^2}$ and $X^{ab}$ is an antisymmetric 2-form accounting for the contribution of the shock wave 
\begin{eqnarray}
& & X^{1a} = (d-1)\, \alpha\,  f(u)\, L^2\, z^{d-1}\; e^0\wedge e^a ~,\qquad a\neq 0,6 ~, \nonumber \\ [0,5em]
& & X^{16} = - [(d-2)^2 - 1]\, \alpha\, f(u)\, L^2\, z^{d-1}\; e^0 \wedge e^6 ~. \nonumber
\end{eqnarray}
Now, the relevant equation is, as before, given by a single component of (\ref{linnear}),
\begin{eqnarray}
\delta \mathcal{E}^{(d)}_3 & \approx & 2\, \epsilon_{{023f_3}\cdots{f_{d-1}}}\, d(\delta \omega^{02}) \wedge \left( 2 R^{f_3f_4} + \frac{1}{\lambda L^2} e^{f_3f_4} \right) \wedge e^{{f_5}\cdots{f_{d-1}}}  \nonumber \\ [0,5em]
& & \qquad +\, 2\, \epsilon_{{123f_3}\cdots{f_{d-1}}}\, d(\delta \omega^{12}) \wedge \left( 2 R^{f_3f_4} + \frac{1}{\lambda L^2} e^{f_3f_4} \right) \wedge e^{{f_5}\cdots{f_{d-1}}} ~. \nonumber
\end{eqnarray}
Using expressions (\ref{dw02}) and (\ref{dw12}) we get
\begin{eqnarray}
\delta \mathcal{E}^{(d)}_3 & \approx &
2 \frac{L^2 z^2}{N_{\#}^2} \left[\left(\partial_u\partial_v\phi+ \alpha f(u) L^2 z^{d-1}\partial_v^2\phi\right)\,\left(2\Lambda+\frac{1}{\lambda L^2}\right)\epsilon_{302f_3 \cdots f_{d-1}} e^{02f_3\cdots{f_{d-1}}} \right.\nonumber\\ [0.5em]
& & \qquad -\, 4\, \partial_v^2\phi\, \Lambda \epsilon_{3021f_4\cdots f_{d-1} }e^{12} \wedge X^{1f_4} \wedge e^{f_5\cdots f_{d-1}}\nonumber\\ [0.5em]
& &\left. \qquad +\,\left(\partial_u\partial_v\phi+ \alpha f(u) L^2 z^{d-1}\partial_v^2\phi\right)\,\left(2\Lambda+\frac{1}{\lambda L^2}\right)\epsilon_{312f_3 \cdots f_{d-1}} e^{12f_3\cdots{f_{d-1}}} \right] ~, \nonumber
\end{eqnarray} 
which after some manipulations can be cast into the form:
\begin{eqnarray}
\delta \mathcal{E}^{(d)}_3 & \approx & 4\frac{L^2 z^2}{N_{\#}^2}\left[(d-3)!\left(\partial_u\partial_v\phi+ \alpha f(u) L^2 z^{d-1}\partial_v^2\phi\right)\left(2\Lambda+\frac{1}{\lambda L^2}\right)\right.\nonumber\\ [0.5em]
& &\left. \quad -\, 2(d-5)!\left((d-5)(d-1)-(d-2)^2+1\right)\alpha f(u)L^2 z^{d-1}\frac{L^2 z^2}{N_{\#}^2}\partial_v^2\phi\,\Lambda\right] ~, \nonumber
\end{eqnarray} 
since the diagonal part of $d(\delta\omega^{ab})$ contributes with all the diagonal parts of $R^{ab}$ and the out-of-diagonal part of $d(\delta\omega^{ab})$ contributes with the out-of-diagonal part of one of the $R^{ab}$. We neglected the overall volume form in the last expression. Collecting terms of the type $\partial_u\partial_v\phi$ and $\partial_v^2\phi$, we obtain $\delta \mathcal{E}^{(d)}_3 = 0$, with
\begin{eqnarray}
\delta \mathcal{E}^{(d)}_3 & \propto & \left[ (d-3)(d-4)\, \partial_u\partial_v\phi \left(1+2\lambda L^2 \Lambda\right) +  \left( (d-3)(d-4)\left(1+2\lambda L^2 \Lambda\right) \right. \right. \nonumber \\ [0.5em]
& & \left. \left. \qquad +\, 4\, (d-1)\lambda L^2 \Lambda \right) \alpha\, f(u)\, L^2 z^{d-1}\, \partial_v^2 \phi \right] \nonumber \\ [0.5em]
& = & \frac{(d-3)(d-4)}{N_{\#}^2} \bigg[(N_{\#}^2-2\lambda)\partial_u\partial_v\phi 
+\, \left(N_{\#}^2-\frac{2(d^2-5d+10)}{(d-3)(d-4)}\,\lambda\right)\alpha f(u) L^2 z^{d-1} \partial_v^2\phi\bigg] \nonumber
\end{eqnarray}
where we have used $L^2 \Lambda=-1/N_{\#}^2$. Causality problems appear when the coefficient of $\partial_v^2\phi$ becomes negative. 
\begin{equation}
N_{\#}^2-\frac{2(d^2-5d+10)}{(d-3)(d-4)}\,\lambda > 0 ~.
\end{equation}
This is the same bound requirement we found using the black hole background. The other two helicities match this expectations as well. The helicity two case gives a negative coefficient for $\lambda$ and therefore an upper bound for it, and the other two cases give positive coefficients and so lower bounds.

A conformal collider {\it gedanken} experiment in a higher dimensional CFT would yield positive energy bounds from (\ref{t2andt4-d}):
\begin{eqnarray}
{\rm tensor:}\qquad & & 1 - \frac{1}{d-2}\, t_2 - \frac{2}{d (d-2)}\, t_4 \geq 0 ~,\label{tensorboundd} \\ [0.7em]
{\rm vector:}\qquad & & \left( 1 - \frac{1}{d-2}\, t_2 - \frac{2}{d (d-2)}\, t_4 \right) + \frac{1}{2}\, t_2 \geq 0 ~, \label{vectorboundd} \\ [0.7em]
{\rm scalar:}\qquad & & \left( 1 - \frac{1}{d-2}\, t_2 - \frac{2}{d (d-2)}\, t_4 \right) + \frac{d-3}{d-2} \left( t_2 + t_4 \right) \geq 0 ~. \label{scalarboundd}
\end{eqnarray}
These restrictions define an allowed triangle in the generic case, as depicted in Figure \ref{triangle}. Assuming $t_4 = 0$,\footnote{This condition is usually valid just for supersymmetric theories. There is a general theorem by Nahm which forbids the existence of SCFTs in space-times of higher dimensions \cite{Nahm1978}. However, it seems possible to overcome the hypothesis of this theorem \cite{Holten1982,Troncoso1998}. We think that this issue deserves further study.} as it is indeed the case in supersymmetric $d=5,7$ theories, this implies, respectively,
\begin{equation}
t_2 \leq d-2 ~, \qquad t_2 \geq -\frac{2(d-2)}{d-4} ~, \qquad t_2 \geq -\frac{d-2}{d-4} ~.
\end{equation}
Once identified the bounds coming from the field theory side, we can try to extract the maximum amount of information matching them with the bounds coming from the gravity side. We can parameterize, in general, the function $t_2(d,\lambda)$ as follows,
\begin{equation}
t_2(d,\lambda) = \frac{a(d)}{\sqrt{1-4\lambda}} + b(d) +  g(d,\lambda) ~.
\label{t2guess}
\end{equation}
Notice that the expressions for the case of 4d and 6d SCFTs (that is, $d = 5, 7$), correspond to $g = 0$ in (\ref{t2guess}). Thus, $g(d,\lambda)$ accounts for any contribution different from those already known by means of holographic Weyl anomaly computations in the lower dimensional cases. Matching the expressions for the bounds coming from both sides, we obtain a set of three algebraic equations; one for each helicity,
\begin{eqnarray}
& & (d^2-5d+10)\; a(d) + 2 (d-1)\; b(d)  = 2 (d-1)(d-2) - 2 (d-1)\; g(d,\lambda^{\star}_2) ~, \nonumber\\ [0.5em]
& & 2\; a(d) + (d-1)\; b(d)  = - \frac{2(d-1)(d-2)}{d-4} - (d-1)\; g(d,\lambda^{\star}_1) ~, \label{system} \\ [0.35em]
& & (d+1)\; a(d) + 2 (d-1)\; b(d)  = - \frac{2(d-1)(d-2)}{d-4} - 2 (d-1)\; g(d,\lambda^{\star}_0) ~, \nonumber
\end{eqnarray}
where $\lambda^{\star}_i(d)$ are the critical values of the Gauss-Bonnet coupling that saturate each of the bounds. We have in principle an over-constrained system of equations for $a(d)$ and $b(d)$. We can, however, solve the three possible pairs of equations for, say, $a(d)$ and then demand that all solutions agree. It is not hard to see that this happens when the three indeterminate functions are equal, $g(d,\lambda^{\star}_i) = g^\star(d)$, for all helicities. We can thus reabsorb this contribution on $b(d)$, which amounts to setting $g^\star(d)=0$. The solution of the system (\ref{system}) simply reads
\begin{equation}
a(d) = -b(d) = \frac{2(d-1)(d-2)}{(d-3)(d-4)} ~.
\end{equation}
Hence, the function $t_2(d,\lambda)$ possibly has no additional contribution $g(d,\lambda)$ unless this function has the quite unnatural property that it vanishes for all the critical values of the coupling, $\lambda_i^\star$, and also, for all values of $\lambda$, in the $d = 5, 7$ cases. The only possibility for such weird extra contribution to $t_2$ would then be $g(d,\lambda) = (d-5) (d-7) (\lambda-\lambda^{\star}_0) (\lambda-\lambda^{\star}_1)(\lambda-\lambda^{\star}_2)\; \tilde{g}(d,\lambda)$, where $\tilde{g}$ is an arbitrary function whose form has to be determined by additional means, and cannot be detected by our analysis. It seems reasonable to assume the absence of such unnatural term and consequently conjecture an expression for $t_2(d,\lambda)$ valid for higher dimensional CFTs with vanishing $t_4$,
\begin{equation}
t_2(d,\lambda) = \frac{2(d-1)(d-2)}{(d-3)(d-4)} \left( \frac{1}{\sqrt{1-4\lambda}} - 1 \right) ~.
\label{t2lambda}
\end{equation}
If there are higher dimensional CFTs with vanishing $t_4$, and the holographic Weyl anomaly computation can be generalized to those theories, (\ref{t2lambda}) should hold.\footnote{Indeed, this expression for $t_2(d,\lambda)$ was confirmed later on by an explicit calculation in \cite{Buchel2009c}, shortly after the original submission of the present article.}
\vskip13mm

%%%%%%%%%%%%%%%%%%%%%%%%%%%%%%%%%%%%%%%%%%%%%%%%%%%%
\section{Discussion}
%%%%%%%%%%%%%%%%%%%%%%%%%%%%%%%%%%%%%%%%%%%%%%%%%%%%

In this article we have computed all polarization linear perturbations of the black hole AdS solution in Gauss-Bonnet theory. We have scrutinized the conditions leading to superluminal propagation in the boundary and thus provided a bound for the GB coupling due to causality in any space-time dimension. In the context of AdS/CFT for 6d SCFTs, our results complement those in \cite{Boer2009} to provide a complete analysis of the correspondence between causality and positive energy conditions emerging in a conformal collider physics setup.

We did the computation following a second approach introduced in \cite{Hofman2009}. We studied the collision of gravitons and shock waves in AdS in the framework of Gauss-Bonnet theory, and found the same constraints on the GB coupling, also for arbitrary space-time dimension. This provides an appealing test of AdS/CFT. The shock wave computation can possibly be argued to be over-constrained by conformal symmetry (since it solely relies on a 3-point function in a CFT), but the computation in the black hole background is less trivially restricted or, at least, the restrictions are far less obvious. It may be the case that the Gauss-Bonnet theory is simple enough so that the 3-point function fully determines the spectrum of boundary perturbations in the black hole background. In this respect, the study of this problem in presence of higher curvature corrections might be clarifying.

The smooth extension to higher dimensions of our results is amusing. The behavior of the upper/lower bound on $\lambda$ is monotonically increasing/decreasing with $d$. In the former case, there is an obstruction in the value $\lambda = 1/4$, where there is symmetry enhancement (a case that should be analyzed separately). It is not surprising, thus, that $\lambda_{\rm max}$ asymptotically approaches $1/4$. It is more striking what happens to the lower bound. There is no a priori restriction. There is no critical negative value of $\lambda$, at least manifestly. Thus, naively one might expect that $\lambda_{\rm min} \to - \infty$ in the infinite dimensional limit. However, the result we obtain is $\lambda_{\rm min} \to - 3/4$. This result, we think, calls for a proper interpretation. In particular, it seems to suggest that $\eta/s$ is order $1/4\pi$ for strongly coupled CFTs in arbitrary dimensions. If true, this may shed further light in generic hydrodynamics features of strongly coupled plasmas.

Another important issue has to do with the string theory origin of the curvature squared corrections discussed in this paper. It seems more natural to seek for the origin of curvature squared corrections to the Einstein-Hilbert Lagrangian in the realm of M-theory. Indeed, the two known maximally $(2,0)$ SCFTs in 6d are the free tensor multiplet theory describing the low energy dynamics of a single M5-brane, and the interacting $(2,0)$ SCFT describing $N$ coincident M5-branes. It is tempting to ask whether this action could be the source of the curvature-squared terms in M-theory. The near extremal limit of the background generated by a stack of M5-branes, indeed, is AdS$_7 \times$ S$^4$. Compactification of 11d supergravity in the S$^4$, though, will generically lead to $R^4$ terms. In presence of probe sources, however, curvature squared corrections can be generated \cite{Bachas1999}. The origin of quadratic curvature corrections in 7d coming from M-theory can be traced to the existence of $A_{k-1}$ singularities that result from a $\IZ_k$ orbifold of S$^4$ \cite{Gaiotto2009d}. The coefficient in front can be obtained by starting from the case of $k$ D6-branes in type IIA string theory and uplifting to M-theory. In the string theory setup, it was recently discussed in \cite{Buchel2009} how curvature squared corrections may arise for the 5d case. The arguments of that paper can be extended to the 7d case smoothly.

We have shown in this paper that the bounds obeyed by the Gauss-Bonnet coupling and those satisfied by the central charges of 4d and 6d SCFTs perfectly match. In the field theory side the bounds are originated in positive energy arguments while in the gravity side it is obtained from causality constraints. We scrutinized the different helicities and showed that the bound originated in each of them satisfies the expectations from AdS/CFT. The higher dimensional results are appealing and call for at least two avenues of research: diving into the yet obscure field theory side or studying the influence of higher curvature corrections to the results of this paper. These seem very interesting problems in which we hope to be able to report soon.

%%%%%%%%%%%%%%%%%%%%%%%%%%%%%%%%%%%%%%%%%%%%%%%%
%% BACKMATTER
%%%%%%%%%%%%%%%%%%%%%%%%%%%%%%%%%%%%%%%%%%%%%%%%

\vskip 15pt
\centerline{\bf Acknowledgments}
\vskip 10pt
\noindent
We would like to thank Alex Buchel, Manuela Kulaxizi, Juan Maldacena, Andrei Parnachev and Jorge Zanelli for interesting comments and insights, and Diego Hofman for collaboration at several stages of this project and lots of discussions. This work is supported in part by MICINN and FEDER (grant FPA2008-01838), by Xunta de Galicia (Conseller\'\i a de Educaci\'on and grant PGIDIT06PXIB206185PR), and by the Spanish Consolider-Ingenio 2010 Programme CPAN (CSD2007-00042). JDE is a {\it Ram\'on y Cajal} Research Fellow and XOC is supported by a spanish FPU fellowship. The Centro de Estudios Cient\'\i ficos (CECS) is funded by the Chilean Government through the Millennium Science Initiative and the Centers of Excellence Base Financing Program of Conicyt. CECS is also supported by a group of private companies which at present includes Antofagasta Minerals, Arauco, Empresas CMPC, Indura, Naviera Ultragas and Telef\'onica del Sur.

%\bibliographystyle{JHEP}
%\bibliography{references}

\begin{thebibliography}{10}

\bibitem{Maldacena1998}
J.~M. Maldacena, {\it {The large N limit of superconformal field theories and
  supergravity}},  {\em Adv. Theor. Math. Phys.} {\bf 2} (1998) 231--252,
  [\href{http://arxiv.org/abs/hep-th/9711200}{{\tt hep-th/9711200}}].

\bibitem{Klebanov2008a}
I.~R. Klebanov, {\it {Testing the AdS/CFT Correspondence}},  {\em AIP Conf.
  Proc.} {\bf 1031} (2008) 3.

\bibitem{Hofman2008}
D.~M. Hofman and J.~Maldacena, {\it {Conformal collider physics: Energy and
  charge correlations}},  {\em JHEP} {\bf 05} (2008) 012,
  [\href{http://arxiv.org/abs/0803.1467}{{\tt arXiv:0803.1467}}].

\bibitem{Basham1978}
C.~L. Basham, L.~S. Brown, S.~D. Ellis, and S.~T. Love, {\it {Electron -
  Positron Annihilation Energy Pattern in Quantum Chromodynamics:
  Asymptotically Free Perturbation Theory}},  {\em Phys. Rev.} {\bf D17} (1978)
  2298.

\bibitem{Basham1978a}
C.~L. Basham, L.~S. Brown, S.~D. Ellis, and S.~T. Love, {\it {Energy
  Correlations in electron - Positron Annihilation: Testing QCD}},  {\em Phys.
  Rev. Lett.} {\bf 41} (1978) 1585.

\bibitem{Birrell}
N.~D. Birrell and P.~C.~W. Davies, {\it Quantum fields in curved space}, .
  Cambridge University Press, UK (1982) 340pp.

\bibitem{Henningson1998}
M.~Henningson and K.~Skenderis, {\it {The holographic Weyl anomaly}},  {\em
  JHEP} {\bf 07} (1998) 023, [\href{http://arxiv.org/abs/hep-th/9806087}{{\tt
  hep-th/9806087}}].

\bibitem{Henningson2000}
M.~Henningson and K.~Skenderis, {\it {Holography and the Weyl anomaly}},  {\em
  Fortsch. Phys.} {\bf 48} (2000) 125--128,
  [\href{http://arxiv.org/abs/hep-th/9812032}{{\tt hep-th/9812032}}].

\bibitem{Brigante2008}
M.~Brigante, H.~Liu, R.~C. Myers, S.~Shenker, and S.~Yaida, {\it {Viscosity
  Bound Violation in Higher Derivative Gravity}},  {\em Phys. Rev.} {\bf D77}
  (2008) 126006, [\href{http://arxiv.org/abs/0712.0805}{{\tt
  arXiv:0712.0805}}].

\bibitem{Brigante2008a}
M.~Brigante, H.~Liu, R.~C. Myers, S.~Shenker, and S.~Yaida, {\it {The Viscosity
  Bound and Causality Violation}},  {\em Phys. Rev. Lett.} {\bf 100} (2008)
  191601, [\href{http://arxiv.org/abs/0802.3318}{{\tt arXiv:0802.3318}}].

\bibitem{Neupane2009a}
I.~P. Neupane and N.~Dadhich, {\it {Entropy bound and causality violation in
  higher curvature gravity}},  {\em Class. Quant. Grav.} {\bf 26} (2009)
  015013.

\bibitem{Neupane2009f}
I.~P. Neupane, {\it {Black Holes, Entropy Bound and Causality Violation}},
  {\em Int. J. Mod. Phys.} {\bf A24} (2009) 3584--3591,
  [\href{http://arxiv.org/abs/0904.4805}{{\tt arXiv:0904.4805}}].

\bibitem{Kats2009}
Y.~Kats and P.~Petrov, {\it {Effect of curvature squared corrections in AdS on
  the viscosity of the dual gauge theory}},  {\em JHEP} {\bf 01} (2009) 044,
  [\href{http://arxiv.org/abs/0712.0743}{{\tt arXiv:0712.0743}}].

\bibitem{Buchel2009}
A.~Buchel, R.~C. Myers, and A.~Sinha, {\it {Beyond eta/s = 1/4pi}},  {\em JHEP}
  {\bf 03} (2009) 084, [\href{http://arxiv.org/abs/0812.2521}{{\tt
  arXiv:0812.2521}}].

\bibitem{HofMalpvt}
D.~M. Hofman and J.~Maldacena, private communication, 2008.

\bibitem{Kovtun2005}
P.~Kovtun, D.~T. Son, and A.~O. Starinets, {\it {Viscosity in strongly
  interacting quantum field theories from black hole physics}},  {\em Phys.
  Rev. Lett.} {\bf 94} (2005) 111601,
  [\href{http://arxiv.org/abs/hep-th/0405231}{{\tt hep-th/0405231}}].

\bibitem{Buchel2009a}
A.~Buchel and R.~C. Myers, {\it {Causality of Holographic Hydrodynamics}},
  \href{http://arxiv.org/abs/0906.2922}{{\tt arXiv:0906.2922}}.

\bibitem{Hofman2009}
D.~M. Hofman, {\it {Higher Derivative Gravity, Causality and Positivity of
  Energy in a UV complete QFT}},  {\em Nucl. Phys.} {\bf B823} (2009) 174--194,
  [\href{http://arxiv.org/abs/0907.1625}{{\tt arXiv:0907.1625}}].

\bibitem{Witten1995c}
E.~Witten, {\it {Some comments on string dynamics}},
  \href{http://arxiv.org/abs/hep-th/9507121}{{\tt hep-th/9507121}}.

\bibitem{Strominger1996d}
A.~Strominger, {\it {Open p-branes}},  {\em Phys. Lett.} {\bf B383} (1996)
  44--47, [\href{http://arxiv.org/abs/hep-th/9512059}{{\tt hep-th/9512059}}].

\bibitem{Boer2009}
J.~de~Boer, M.~Kulaxizi, and A.~Parnachev, {\it {AdS$_7$/CFT$_6$, Gauss-Bonnet
  Gravity, and Viscosity Bound}},  \href{http://arxiv.org/abs/0910.5347}{{\tt
  arXiv:0910.5347}}.

\bibitem{Witten2007c}
E.~Witten, {\it {Conformal Field Theory In Four And Six Dimensions}},
  \href{http://arxiv.org/abs/0712.0157}{{\tt arXiv:0712.0157}}.

\bibitem{Bachas1999}
C.~P. Bachas, P.~Bain, and M.~B. Green, {\it {Curvature terms in D-brane
  actions and their M-theory origin}},  {\em JHEP} {\bf 05} (1999) 011,
  [\href{http://arxiv.org/abs/hep-th/9903210}{{\tt hep-th/9903210}}].

\bibitem{Gaiotto2009d}
D.~Gaiotto and J.~Maldacena, {\it {The gravity duals of N=2 superconformal
  field theories}},  \href{http://arxiv.org/abs/0904.4466}{{\tt
  arXiv:0904.4466}}.

\bibitem{Edelstein2001}
J.~D. Edelstein and C.~Nunez, {\it {D6 branes and M-theory geometrical
  transitions from gauged supergravity}},  {\em JHEP} {\bf 04} (2001) 028,
  [\href{http://arxiv.org/abs/hep-th/0103167}{{\tt hep-th/0103167}}].

\bibitem{Osborn1994}
H.~Osborn and A.~C. Petkou, {\it {Implications of Conformal Invariance in Field
  Theories for General Dimensions}},  {\em Ann. Phys.} {\bf 231} (1994)
  311--362, [\href{http://arxiv.org/abs/hep-th/9307010}{{\tt hep-th/9307010}}].

\bibitem{Bonora1986}
L.~Bonora, P.~Pasti, and M.~Bregola, {\it {Weyl Cocycles}},  {\em Class. Quant.
  Grav.} {\bf 3} (1986) 635.

\bibitem{Lovelock1971}
D.~Lovelock, {\it {The Einstein tensor and its generalizations}},  {\em J.
  Math. Phys.} {\bf 12} (1971) 498--501.

\bibitem{Chamseddine1989e}
A.~H. Chamseddine, {\it {Topological gauge theory of gravity in five-dimensions
  and all odd dimensions}},  {\em Phys. Lett.} {\bf B233} (1989) 291.

\bibitem{Zanelli2005}
J.~Zanelli, {\it {Lecture notes on Chern-Simons (super-)gravities}},
  \href{http://arxiv.org/abs/hep-th/0502193}{{\tt hep-th/0502193}}.

\bibitem{Boulware1985a}
D.~G. Boulware and S.~Deser, {\it {String Generated Gravity Models}},  {\em
  Phys. Rev. Lett.} {\bf 55} (1985) 2656.

\bibitem{Cai2002}
R.-G. Cai, {\it {Gauss-Bonnet black holes in AdS spaces}},  {\em Phys. Rev.}
  {\bf D65} (2002) 084014, [\href{http://arxiv.org/abs/hep-th/0109133}{{\tt
  hep-th/0109133}}].

\bibitem{Myers1988}
R.~C. Myers and J.~Z. Simon, {\it {Black Hole Thermodynamics in Lovelock
  Gravity}},  {\em Phys. Rev.} {\bf D38} (1988) 2434--2444.

\bibitem{Nojiri2001j}
S.~Nojiri and S.~D. Odintsov, {\it {Anti-de Sitter black hole thermodynamics in
  higher derivative gravity and new confining-deconfining phases in dual CFT}},
   {\em Phys. Lett.} {\bf B521} (2001) 87--95,
  [\href{http://arxiv.org/abs/hep-th/0109122}{{\tt hep-th/0109122}}].

\bibitem{Cvetic2002}
M.~Cvetic, S.~Nojiri, and S.~D. Odintsov, {\it {Black hole thermodynamics and
  negative entropy in deSitter and anti-deSitter Einstein-Gauss-Bonnet
  gravity}},  {\em Nucl. Phys.} {\bf B628} (2002) 295--330,
  [\href{http://arxiv.org/abs/hep-th/0112045}{{\tt hep-th/0112045}}].

\bibitem{Kofinas2006}
G.~Kofinas and R.~Olea, {\it {Vacuum energy in Einstein-Gauss-Bonnet AdS
  gravity}},  {\em Phys. Rev.} {\bf D74} (2006) 084035,
  [\href{http://arxiv.org/abs/hep-th/0606253}{{\tt hep-th/0606253}}].

\bibitem{Banados1994}
M.~Ba\~nados, C.~Teitelboim, and J.~Zanelli, {\it {Dimensionally continued
  black holes}},  {\em Phys. Rev.} {\bf D49} (1994) 975--986,
  [\href{http://arxiv.org/abs/gr-qc/9307033}{{\tt gr-qc/9307033}}].

\bibitem{Garraffo2008}
C.~Garraffo and G.~Giribet, {\it {The Lovelock Black Holes}},  {\em Mod. Phys.
  Lett.} {\bf A23} (2008) 1801--1818,
  [\href{http://arxiv.org/abs/0805.3575}{{\tt arXiv:0805.3575}}].

\bibitem{Horowitz1999a}
G.~T. Horowitz and N.~Itzhaki, {\it {Black holes, shock waves, and causality in
  the AdS/CFT correspondence}},  {\em JHEP} {\bf 02} (1999) 010,
  [\href{http://arxiv.org/abs/hep-th/9901012}{{\tt hep-th/9901012}}].

\bibitem{Ge2009}
X.-H. Ge and S.-J. Sin, {\it {Shear viscosity, instability and the upper bound
  of the Gauss-Bonnet coupling constant}},  {\em JHEP} {\bf 05} (2009) 051,
  [\href{http://arxiv.org/abs/0903.2527}{{\tt arXiv:0903.2527}}].

\bibitem{Nahm1978}
W.~Nahm, {\it {Supersymmetries and their representations}},  {\em Nucl. Phys.}
  {\bf B135} (1978) 149.

\bibitem{Holten1982}
J.~W. van Holten and A.~Van~Proeyen, {\it {N=1 Supersymmetry Algebras in D=2,
  D=3, D=4 mod-8}},  {\em J. Phys.} {\bf A15} (1982) 3763.

\bibitem{Troncoso1998}
R.~Troncoso and J.~Zanelli, {\it {New gauge supergravity in seven and eleven
  dimensions}},  {\em Phys. Rev.} {\bf D58} (1998) 101703,
  [\href{http://arxiv.org/abs/hep-th/9710180}{{\tt hep-th/9710180}}].

\bibitem{Buchel2009c}
A.~Buchel, J.~Escobedo, R.~C. Myers, M.~F. Paulos, A.~Sinha, and M.~Smolkin,
  {\it {Holographic GB gravity in arbitrary dimensions}},
  \href{http://arxiv.org/abs/0911.4257}{{\tt arXiv:0911.4257}}.

\end{thebibliography}

\providecommand{\href}[2]{#2}\begingroup\raggedright\endgroup
\end{document}